\documentclass[a4paper,11pt]{article}
\pdfoutput=1 

\usepackage{jcappub} 

\usepackage[T1]{fontenc} 
\usepackage{epstopdf}
\usepackage{mathtools}
\usepackage{graphicx}

\title{Gravitational lensing by a stable rotating regular black hole}

\author{Chen-Hao Xie,}
\author[1]{Yu Zhang,\note{Corresponding author.}}
\author{Qi Sun,}
\author{Qi-Quan Li}
\author{and Peng-Fei Duan}

\affiliation{Faculty of Science, Kunming University of Science and Technology,\\Kunming 650500, China}

\emailAdd{chenhaoxie2000@163.com}
\emailAdd{zhangyu\_128@126.com}
\emailAdd{549251942@qq.com}
\emailAdd{1015852575@qq.com}
\emailAdd{dpffqs@mail.ustc.edu.cn}

\abstract{Recent observational data from the Event Horizon Telescope (EHT) collaboration provide convincing realistic evidence for the existence of black hole rotation. From a phenomenological perspective, a recently proposed stable rotating regular (SRR) black hole circumvents the theoretical flaws of the Kerr solution. For the purpose of obtaining observational signatures of this black hole, we study its gravitational lensing effect. In the strong field limit, we calculate the deflection angle of light, the radius of the photon sphere, and other observables. The observables include the relativistic image position, separation, magnification, and time delays between different images. Then, by modeling M87* and Sgr A* as the SRR black hole, we compute their observables and evaluate the deviation of the observables from the Kerr case. In the weak field limit, we calculate the light deflection angle of M87* and Sgr A* via the Gauss-Bonnet theorem (GBT). With the growth of deviation parameter $e$, the gravitational lensing effect in the weak field limit intensifies monotonically, and the gravitational lensing effect in the strong field limit changes dramatically only at high spins. Our research may contribute to distinguish between SRR black holes from Kerr black holes under higher-precision astronomical observations.}


\makeatletter
\gdef\@fpheader{}
\makeatother
\begin{document}
\maketitle
\flushbottom

\section{Introduction}
\label{sec:intro}

As one of the exact solutions to Einstein field equations, the Kerr metric describes the structure of spacetime about a rotating, axially symmetric black hole. The recent studies \cite{EventHorizonTelescope:2019dse,EventHorizonTelescope:2019ggy,EventHorizonTelescope:2019pgp,EventHorizonTelescope:2022wkp,EventHorizonTelescope:2022xqj,Cui:2023uyb} strongly support the presence of spinning black holes in M87 and Sgr A. However, Kerr spacetime still has three main flaws: the scalar curvature caused by the central singularity tends to infinity \cite{Penrose:1964wq,Hawking:1973uf,Wald:1984rg}, the instability results from the mass inflation \cite{Poisson:1989zz} of the Cauchy horizon, and the presence of closed time-like curves (CTCs). In order to solve the problem of the central singularity, Franzin et al. \cite{Franzin:2022wai} constructed the Kerr metric as a solution of theories \cite{Englert:1976ep,Narlikar:1977nf,Bars:2013yba,DominisPrester:2013rpi} having conformal symmetry by multiplying the metric with a conformal factor. Simultaneously, the conformal factor prevents particles reaching the center of the black hole, thus avoiding CTCs \cite{Bambi:2016yne}. Furthermore, the authors used the mass function to stabilize mass inflation, and finally obtained a solution of SRR black hole. The effective stress-energy tensor that describes the distribution and properties of matter inside a black hole may locally violate some energy conditions \cite{Franzin:2022wai}. These violations might be manifestations of quantum gravitational effects, which can be explained by some higher-order gravitational actions related to the low-energy limit of the quantum gravity theory.

Black holes are often used to interpret astronomical observations. Essentially, they are the best laboratories to test strong gravity. In these laboratories, strong and weak gravitational lensing \cite{Bartelmann:1999yn,Perlick:2004tq,Bozza:2010xqn} serve as powerful astrophysical tools \cite{Aubourg:1993wb} for studying the characteristics of strong gravitational fields and detecting dark matter \cite{Clowe:2006eq}, respectively. Gravitational lensing is the phenomenon of the bending of light by matter. The angle of light deflection depends on the properties of the lens itself and the distance of the observer from the lens. This demonstrates that the gravitational lensing effect of black holes can reveal information about their nature, thereby allowing us to distinguish between different black holes \cite{Eiroa:2002mk,Gyulchev:2006zg,Bin-Nun:2009hct,Kraniotis:2010gx,Wei:2011nj,Kraniotis:2014paa,Zhao:2016kft,Chakraborty:2016lxo,Zhao:2017cwk,Jusufi:2018jof,Wang:2019cuf,Jusufi:2019caq,Kumar:2020sag,Ghosh:2020spb} and test various gravity theories \cite{Bhadra:2003zs,Horvath:2011xr,Sahu:2015dea,Badia:2017art,Allahyari:2019jqz,Li:2020zxi,Afrin:2021imp,Afrin:2021wlj,Kuang:2022ojj,Kuang:2022xjp,Vagnozzi:2022moj,Pantig:2022ely,Soares:2023err}.

When a large number of photons pass nearby, black holes will reveal a dark shadow, the photon ring, and the relativistic images caused by the gravitational lensing effect at the event horizon. Since the black hole shadow resulting from gravitational lensing effect was proposed, related research has been booming \cite{Falcke:1999pj,Takahashi:2004xh,Bambi:2008jg,Hioki:2009na,Amarilla:2011fx,Grenzebach:2014fha,Cunha:2015yba,Abdujabbarov:2015pqp,Abdujabbarov:2016hnw,Younsi:2016azx,Tsukamoto:2017fxq,Cunha:2018acu,Shaikh:2018lcc,Shaikh:2018kfv,Gralla:2019xty,Vagnozzi:2019apd,Konoplya:2019fpy,Konoplya:2021slg,Ghosh:2022gka}. This significant gravitational lensing effect can be described under the strong field limit. Research into the strong field limit can be traced back to the work of Darwin \cite{Darwin:1959wai}. After that, Virbhadra and Ellis obtained the gravitational lens equation in the strong field limit and analyzed the gravitational lens of the Schwarzschild black hole \cite{Virbhadra:1999nm}. Later, they explored the naked singularity lens \cite{Virbhadra:2002ju} and relativistic images of Schwarzschild black hole lensing \cite{Virbhadra:2008ws,Virbhadra:2022iiy,Virbhadra:2022ybp}. Then, Bozza et al. proposed an analytical method to calculate the strong deflection angle of a spherically symmetric black hole \cite{Bozza:2002zj}, and calculated the observables of a spherically symmetric black hole based on the lens equation. Moreover, they extended the method to the situation of rotation \cite{Bozza:2002af} and studied the time delay \cite{Bozza:2003cp} between different relativistic images formed by the gravitational lensing of black holes in the strong field limit. Since then, this method has been widely used in the studies of strong deflection of light \cite{Whisker:2004gq,Chen:2009eu,Liu:2010wh,Ding:2010dc,Sotani:2015ewa,Tsukamoto:2016qro,Hsieh:2021scb,Ghosh:2022mka,Tsukamoto:2022tmm,Tsukamoto:2022uoz,AbhishekChowdhuri:2023ekr,Soares:2023uup} and time delay \cite{Eiroa:2013nra,Lu:2016gsf,Cavalcanti:2016mbe,Islam:2021ful,Hsieh:2021rru}. When the distance between the photons and the black hole is large, the gravitational lensing effect of the black hole should be depicted by the weak field limit. Gibbons and Werner pioneered the study of applying the GBT to the optical metric of the lens, and computed the deflection angle of light at the weak field limit of spherically symmetric black hole spacetime \cite{Gibbons:2008rj}. Werner used the osculating Riemann approach in Finsler geometry to discuss the lensing by the Kerr black hole subsequently \cite{Werner:2012rc}. However, it is unlikely that Finsler geometry can be used to calculate finite distance corrections. In order to test the finite distance correction of the deflection angle of light in axially symmetric space time, Ono et al. calculated the weak deflection angle of the Kerr black hole in spatial geometry \cite{Ono:2017pie}.

The purpose of this paper is to explore the effects of the deviation parameter $e$ and the spin parameter $a$ on the strong and weak field limit of gravitational lensing by the SRR black hole with the methods of Bozza \cite{Bozza:2002af} and GBT \cite{Ono:2017pie}, respectively. The paper is structured as follows. Section~\ref{Spacetime} briefly reviews the SRR black hole solution. In section \ref{Strong gravitational lensing}, we study the gravitational lensing effect in the strong field limit and calculate the strong deflection angle. Section~\ref{Observables} models M87* and Sgr A* as a SRR black hole to calculate the strong lensing observables and evaluates their deviation from the Kerr situation. The formalism for gravitational deflection of light in the weak field limit is set up in section~\ref{Weak gravitational lensing}. Finally, we conclude and discuss our results in section~\ref{Conclusion}.

\section{Spacetime}
\label{Spacetime}
The metric of the SRR black hole can be expressed as
\begin{equation}\label{metric}
            d s^{2}=\frac{\Psi }{\Sigma }  \left [ -(1-\frac{2 m(r) r}{\Sigma } ) d t^{2}- \frac{4 a m(r) r \sin^{2}  {\theta }}{\Sigma }  d t d \phi + \frac{\Sigma}{\Delta } d r^{2} + \Sigma  d \theta ^{2} + \frac{A \sin^{2}  {\theta }}{\Sigma }d \phi ^{2} \right ] ,
\end{equation}
where 
\begin{align}
    \Psi (r,\theta ) &= \Sigma +\frac{l}{r^{3}}, \\
    \Sigma &= r^2+a^2\cos^2\theta, \\
    \Delta &= r^2-2m(r)r+a^2, \\
    A &= (r^2+a^2)^2-\Delta a^2\sin^2\theta.
\end{align}

For the singularities issue, conformal theory considers the singularities as an artifact of the conformal gauge, and they can be eliminated by appropriate gauge transformations \cite{Bambi:2016wdn}. Here, the gauge transformation is the rescaling that we perform: Adopting the conformal factor $\Psi (r,\theta )=\Sigma +\frac{l}{r^{3}}$ rescales the metric to make the Ricci scalar and Kretschmann scalar finite everywhere. Namely, the rescaling of the metric regularizes the spacetime singularity. Meantime, the inherent time it takes for a photon to reach the would-be singularity from $ \theta = \frac{\pi}{2} $ and $ \theta = 0 $ is infinite (meaning such motion could not occur), thus avoiding CTCs.

After that, we will adopt the same mass function, internal and external horizons as ref. \cite{Franzin:2022wai} to avoid the instability caused by mass inflation:
\begin{align}
    m(r) &= M\frac{r^{2} \,+\, \alpha r\,+\, \beta }{r^{2} \,+\, \gamma r \,+\, \mu }, \\
    r_{+} &= M + \sqrt{M^2 - a^2}, \\
    r_{-} &= \frac{a^2}{M + (1-e) \sqrt{M^2 - a^2} }.
\end{align}

Among them, the parameter $e$ controls the difference between the inner horizon and the Kerr horizon. The further requirement that the mass function has no poles implies
\begin{equation}
           -3 - \frac{3M}{\sqrt{M^2 - a^2} } < e <2.
\end{equation}

At $ e = 0 $, it returns to the conformal Kerr black hole. When $e \to 2$, it is a limit similar to $a \to M$.

Then, choosing $\beta \ne 0 $ and $\alpha \ne \gamma$, we can remove mass inflation singularity by making the inner horizon degenerate and the surface gravity zero there. The coefficients $\alpha$, $\beta$, $\gamma$ and $\mu$ of mass function are all functions of inner horizon $ r_{-}$ and external horizon $r_{+}$, which are given by \cite{Franzin:2022wai}
\begin{align}
			\alpha &= \frac{a^4 + r_{-}^3 r_{+} - 3 a^2 r_{-} (r_{-} + r_{+})}{2 a^2 M} ,\\
            \beta &= \frac{a^2 (2 M - 3 r_{-} - r_{+}) +  r_{-}^2 (r_{-} + 3r_{+})}{2 M} ,\\
            \gamma &= 2 M - 3 r_{-} - r_{+} ,\\
            \mu &= \frac{r_{-}^3 r_{+}}{a^2} .
\end{align}

For energy conditions, the authors in ref. \cite{Franzin:2022wai} concluded that not only null but also weak and dominant energy conditions are met if $ \alpha < \gamma $.

As mentioned above, the conformal factor $\Psi$ is used to regularize the singularities, and the mass function $m(r)$ is adopted to avoid the mass inflation. Simultaneously, the choice of conformal factor $\Psi$ ensures the non-existence of CTCs.

To sum up, this metric represents a set of stable, rotating, regular spacetimes without CTCs. It has two free parameters (except the commonly used spin parameter $a$): the ``Kerr-deviation parameter" $e$ and the ``regularization parameter" $l$. It's important to note that the metric becomes conformal to the extremal Kerr black hole when $a \to M$, and it is conformal to the Schwarzschild black hole in the case of $a \to 0$.

\section{Strong gravitational lensing}
\label{Strong gravitational lensing}
 In this section, we consider the strong gravitational deflection of light on the equatorial plane ($\theta = \pi / 2$) and in the units of $2M$ with $M = 1/2$ under the spacetime of the SRR black hole. The reduced metric has the form:
\begin{equation}\label{simmetric}
            d s^{2}=-A(x) d t^{2} + B(x) d r^{2} + C(x)d \phi ^{2} - D(x) d t d \phi.
\end{equation}

The Lagrangian of photons writes as $\mathcal{L} =  - \frac{1}{2} g_{\mu \nu} \dot{x^\mu}  \dot{x^\nu} = 0$, where the overdot indicates the derivative with respect to affine parameter $\lambda$ of geodesics. Because the metric of the SRR black hole is stable and axially symmetrical (it is not a function of time $t$ and rotation angle $\phi$), there are two conserved quantities, i.e., energy and angular momentum:
\begin{align}
            E &= \frac{\partial \mathcal{L}}{ \partial \dot t } = - g_{tt} \dot{t} - g_{t\phi} \dot{\phi}, \\
            L &= - \frac{\partial \mathcal{L}}{ \partial \dot \phi } = g_{t \phi} \dot{t} + g_{\phi \phi} \dot{\phi}.
\end{align}

We set $E = 1$ by choosing a suitable affine parameter, and then replace $L / E$ with impact parameter $b$. From these relationships and equations, we get the motion equations of photons:
\begin{align}
           \dot t &= \frac{4 C  - 2 D b}{4 A C + D^2}, \label{motion1}   \\
           \dot r &=  \pm 2 \sqrt{ \frac{C - D b - Ab^2}{ B (4 A C + D^2)} }, \label{motion2} \\
           \dot \phi &= \frac{2 D + 4 A b}{4 A C + D^2}. \label{motion3}
\end{align}

It is worth noting that the parameter $l$ of conformal factor does not affect the movement of photons according to eqs.~(\ref{motion1}),~(\ref{motion2}) and~(\ref{motion3}). In addition, we define the effective potential for the radial motion as
\begin{equation}
           V_{eff} \coloneqq  -\dot r = - \frac{4(C - D b - Ab^2)}{ B (4 A C + D^2)},
\end{equation}
which describes the different kinds of possible orbits.

We can identify a light ray originating from infinity and incident on the equatorial plane by impact parameter. The impact parameter $b$ can be uniquely determined by the closest approach distance $r_0$ and vice versa. When $r = r_0$, the radial velocity of the photon is zero, that is, the effective potential is equal to zero. The relationship between $b$ and $r_0$ can be obtained by solving the following equation:
\begin{equation}\label{pse1}
           b(r_0) = \frac{-D_0 + \sqrt{4 A_0 C_0 + {D_0}^2 }}{2 A_0},
\end{equation}
where the subscript $0$ indicates that the function is evaluated at $r_0$. The deflection angle increases with decrease in the impact parameter $u_0$ (in turn reduces $r_0$). At a critical point, the deflection angle surpasses $2 \pi$. This results in the formation of a complete ring which is known as the photon sphere. These orbits are unstable against small radial perturbations, which would finally drive photons into the black hole or toward infinity. At the critical point, the effective potential reaches the maximum value when
\begin{equation}\label{pse2}
           \frac{d V_{eff}}{d x} \mid_{r_0 = r_m} = 0.
\end{equation}

By combining eqs.~(\ref{pse1}) and~(\ref{pse2}), we can get the photon sphere equation,
 \begin{equation}\label{pse3}
           A C' -A' C + b (A' D - A D') = 0,
\end{equation}
and photon sphere radius $r_m$ is the largest root of the photon sphere equation.

The deflection angle of the light ray is the angle between the incident and exiting trajectories at the closest distance approach $r_0$ which reads as
\begin{align}\label{Intergral}
            \alpha_{Sd}(r_0) &= I(r_0) - \pi, \\
            I(r_0) &= 2 \int_{r_0}^{\infty} \frac{d \phi} {dr} dr.
\end{align}

Following the idea of strong field limit \cite{Bozza:2002zj}, we expect to get an analytical expansion of the deflection angle which diverges close to $r_m$ \cite{Bozza:2002zj,Bozza:2002af,Tsukamoto:2016jzh}:
\begin{equation}\label{seqs3}
            \alpha_{Sd} (\theta) = -\bar{a} \log_{}{\left ( \frac{\theta D_{OL}}{b_m}-1\right)} + \bar{b} + \mathcal{O}((b-b_m) \log(b-b_m)),
\end{equation}
where $\alpha_{Sd} (\theta)$ is defined as the deflection angle at the strong field limit \cite{Bozza:2002zj}. $\bar{a}$ and $\bar{b} $ are known as strong field limit coefficients. Then, we define the variable $z = \frac{A-A_0}{1-A_0}$, and the eq.~(\ref{Intergral}) becomes
\begin{align}
            I(r_0) &= \int_{0}^{1} R(z , r_0) f(z , r_0) dz, \\
            R(z , r_0) &= 2 \frac{(1 - A_0)}{A'} \frac{\sqrt{B}(2 A_0 A b + A_0 D)}{\sqrt{A_0 C} \sqrt{4 A C + D^2}}, \\
            f(z , r_0) &= \frac{1}{\sqrt{A_0 - A \frac{C_0}{C} + \frac{b}{C}(A D_0 - A_0 D)}}.
\end{align}

The function $R(z , r_0)$ is regular everywhere, while $f(z , r_0)$ diverges as $z \to 0$. To determine the order of divergence of the integrand, we expand the argument of the square root in $f(z , r_0)$ to the second order in $z$:
\begin{align}
            f(z,r_0) \sim f_0(z , r_0) = \frac{1}{\sqrt{\bar{\alpha} z + \bar{\beta} z^2}},
\end{align}
where $\bar{\alpha}$ and $\bar{\beta}$ are the coefficients of Taylor series expansion.

Using above relationships, we can obtain the strong field limit coefficients $\bar{a}$ and $ \bar{b}$ in eq.~(\ref{seqs3}) as
\begin{align}
            \bar{a} &= \frac{R(0 , r_m)}{2 \sqrt{\bar{\beta}_m}}, \\
            \bar{b} &= -\pi + \bar{b}_D + \bar{b}_R + \bar{a} \log{ \frac{c r_{m}^2}{b_m}},
\end{align}
where
\begin{align}
            \bar{b}_D &= 2 \bar{a} \log{\frac{2(1 - A_m)}{A'_m r_m}}, \\
            \bar{b}_R &= \int_{0}^{1} \left [   R(z , r_m) f(z , r_m) - R(0 , r_m) f_0(z , r_m)\right ]dz,
\end{align}
and $c$ is defined by
\begin{equation}
            b - b_m = c(r_0 - r_m)^2,
\end{equation}
in which functions with the subscript $m$ are evaluated at $r_0  = r_m$.

\begin{figure}[t]
\centering
\includegraphics[width=0.45\linewidth]{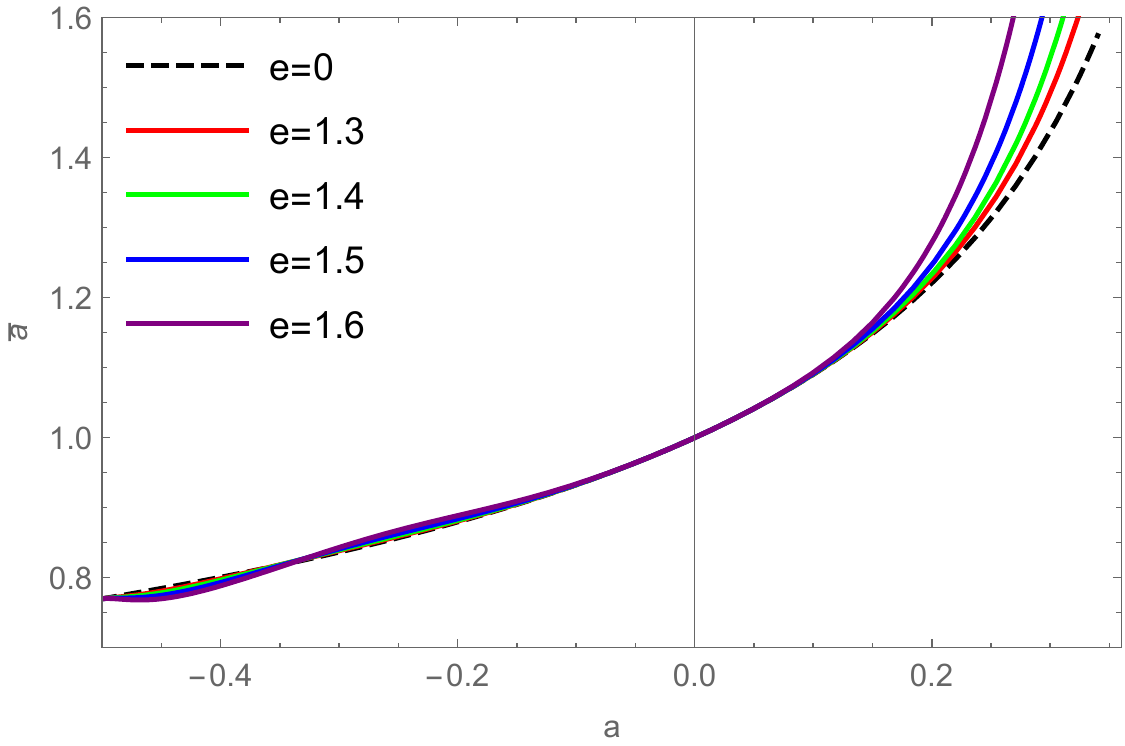}
\includegraphics[width=0.466\linewidth]{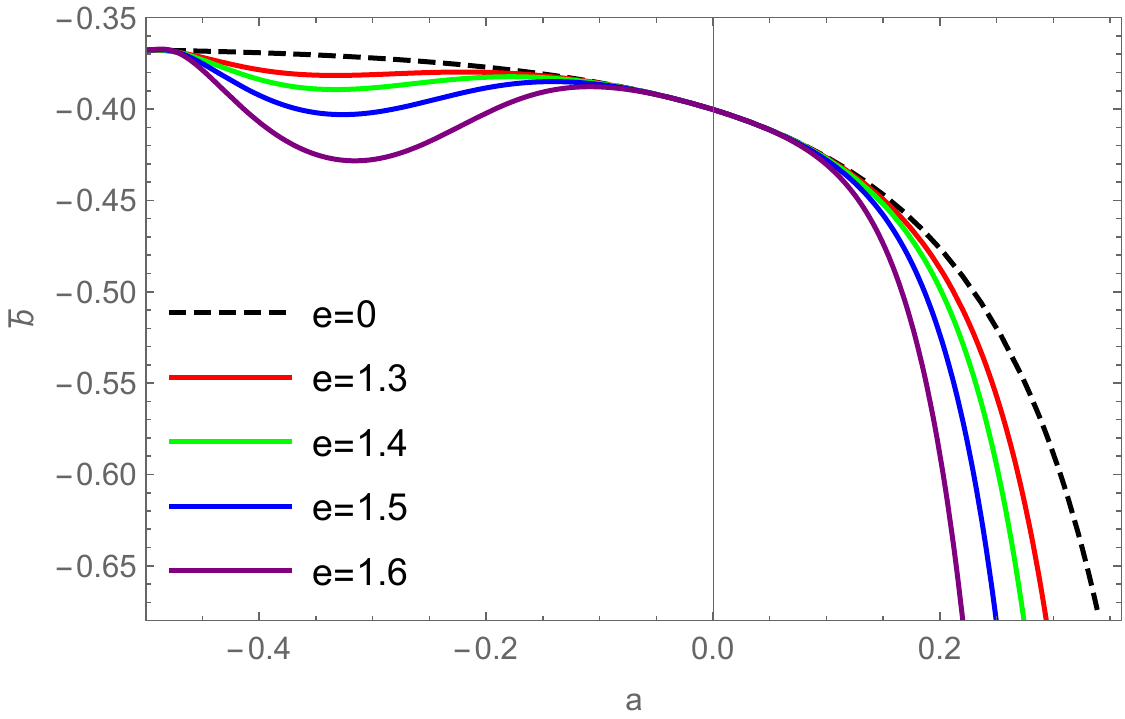}
\caption{The behaviors of the strong lensing coefficients $\bar{a}$ and $\bar{b}$ as function of $a$ for the SRR black hole with different values of $e$.}
\label{sd1a2}
\end{figure}

Now, we will study the deflection angle of the SRR black hole in strong field limit and evaluate the difference from Kerr case. The strong field limit coefficients $\bar{a}$ and $\bar{b}$ are plotted in figure~\ref{sd1a2} to see how they change with the spin parameter $a$ and the deviation parameter $e$. In a slow rotating scenario, we can see that $e$ hardly causes $\bar{a}$ and $ \bar{b}$ of the SRR black hole to diverge significantly from the Kerr situation.

In a high spin scenario for prograde photons ($a>0$), $\bar{a}$ shows an increasing trend and $\bar{b}$ shows a decreasing trend with growth of $a$. And when $a$ is fixed, it is noted that $\bar{a}$ increases and $\bar{b}$ decreases with an increase in $e$. For retrograde photons ($a<0$), with increasing $\left | a \right |$, the deviation parameter $e$ leads to a fluctuation in the strong field limit coefficient $\bar{b}$, which deviates and then converges to the Kerr situation. Simultaneously, $\bar{a}$ generates a double fluctuation that deviates and then converges to the Kerr situation. In addition, the left panel of figure~\ref{sd1a2} shows the growth of the fluctuation amplitude when we increase $e$.

\begin{figure}[t]
\centering
\includegraphics[width=0.453\linewidth]{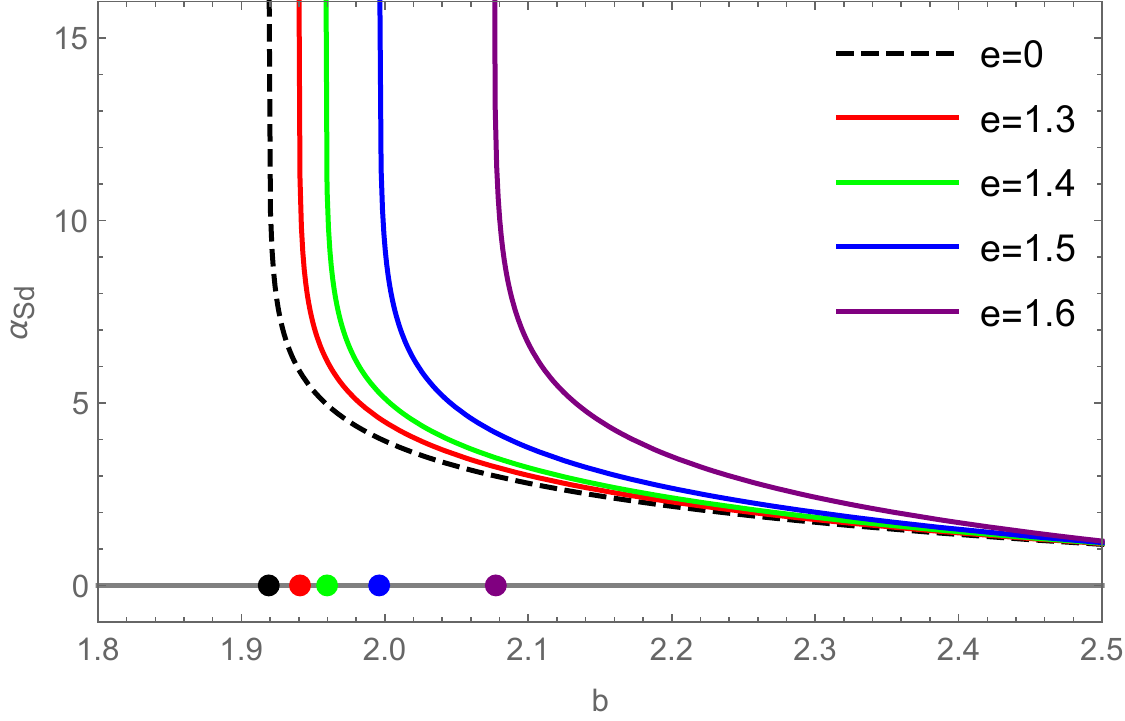}
\includegraphics[width=0.45\linewidth]{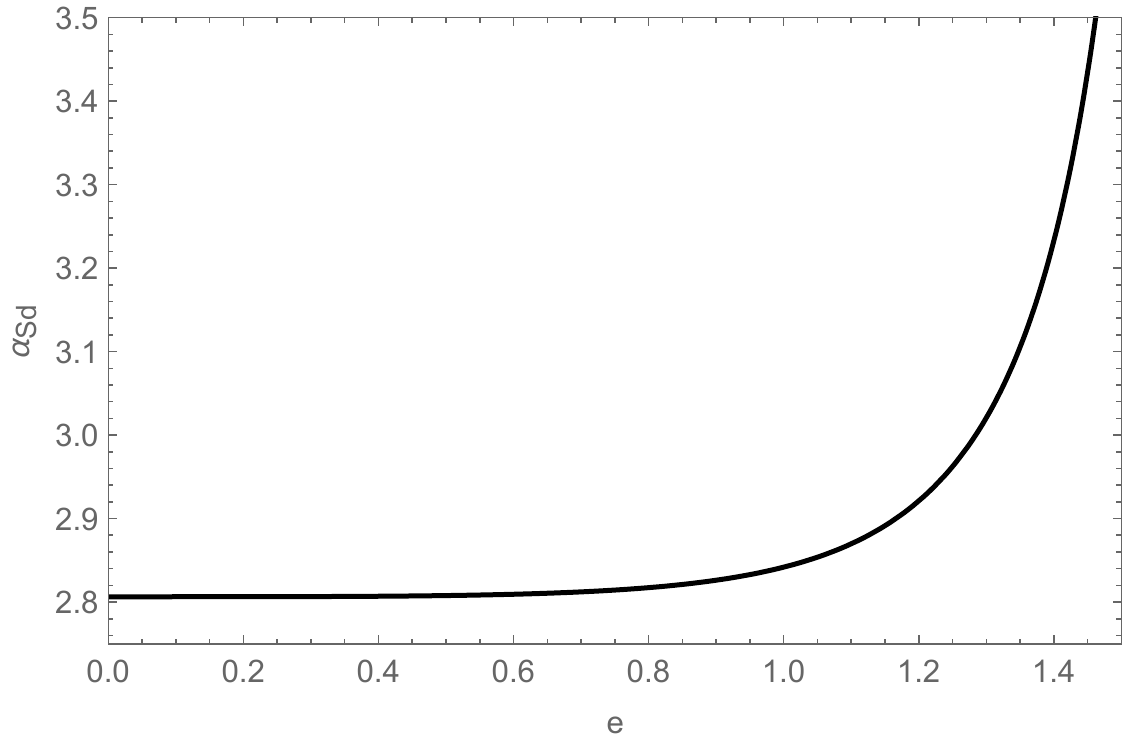}
\caption{Left Panel: The behaviors of the strong deflection $\alpha_{Sd}$ as function of impact parameter $b$ for $a/M = 0.6$ with different values of $e$. The colored points on the horizontal axis correspond to the impact parameter $b = b_m$ for deflection angle diverging. Right Panel: The strong deflection $\alpha_{Sd}$ as function of impact parameter $e$ for $a/M = 0.6$ and $b/M = 2.1$.}
\label{sd3a4}
\end{figure}

We plot the deflection angle $\alpha_{Sd}(\theta)$ in figure~\ref{sd3a4} for different $a$ and $e$. The left panel shows that the deflection angle $\alpha_{Sd}(\theta)$ magnifies monotonically with the reduction of $b$, and diverges at $b =  b_m$. It is obvious that the existence of the deviation parameter $e$ increases $\alpha_{Sd}(\theta)$ and $b_m$. In the right panel of figure~\ref{sd3a4}, we can see that the deflection angle $\alpha_{Sd}(\theta)$ increases with the deviation parameter $e$. Since the strong field limit is only valid when $b$ is close to $b_m$, we cannot obtain a valid result when $b$ is much larger than $b_m$. Therefore, there is a $b = b_z$ making $\alpha_{Sd} = 0$ when $b$ gradually increases.

\section{Observables}
\label{Observables}
The best geometry for observing relativistic images is the source almost aligning with the lens. We can assume that both the source and the observer are far away from the lens, as well as perfectly aligned in asymptotically flat spacetime. Then the lens equation \cite{Virbhadra:1999nm} can be simplified as \cite{Bozza:2001xd,Bozza:2008ev}
\begin{equation} \label{lenseq}
            \beta_{LS}=\theta-\frac{D_{LS}}{D_{OS}}\Delta\alpha_{n},
\end{equation}
where $\Delta\alpha_{n}=\alpha_{Sd}(\theta) - 2n\pi $ is the offset of the deflection angle. $D_{LS}$ and $D_{OL}$ are the distance from the lens to the source and the distance from the observer to the lens, respectively. $\beta_{LS}$ is the angular separation between the lens and the source, and $\theta$ is the angular separation between the lens and the image.

Expanding the offset $\Delta\alpha_{n}$ around $\theta = \theta^0_n$ and letting $\Delta\theta_n = \theta - \theta^0_n $, we find \cite{Bozza:2002zj}
\begin{equation} \label{offset}
            \Delta\alpha_{n}=-\frac{\overline{a}D_{OL}}{b_{m}e_{n}}\Delta\theta_{n}.
\end{equation}

According to eqs.~(\ref{lenseq}) and ~(\ref{offset}), we approximately get the position of the $n^{th}$ relativistic image \cite{Bozza:2002zj}:
\begin{equation}\label{ri}
            \theta_{n} \simeq \theta_{n}^{0}+\frac{b_{m}e_{n}(\beta_{LS}-\theta_{n}^{0})D_{OS}}{\bar{a}D_{LS}D_{OL}},
\end{equation}
where $\theta^0_n$ is the image position corresponding to $\alpha(\theta^0_n) = 2n\pi$ and $e_n=\exp\left(\frac{\bar{b}-2n\pi}{\bar{a}}\right)$.

Due to the effect of gravitational lensing, the brightness of the source image will be magnified. The magnification of the $n^{th}$ relativistic image is defined as \cite{Bozza:2002zj}

\begin{equation}\label{magnification}
            \mu_{n}=\left(\frac{\beta_{LS}}{\theta}\frac{d\beta_{LS}}{d\theta}\right)^{-1}\Bigg|_{\theta_{n}^{0}}=\frac{b_{m}^{2}e_{n}\big(1+e_{n}\big)D_{OS}}{\bar{a}\beta_{LS} D_{LS}D_{OL}^{2}},
\end{equation}
where $\mu_{n}$ decays very quickly with $n$, so the first relativistic image is the brightest. Moreover, it is proportional to a very small value of $1/D^2_{OL}$, so that we can get a bright image only when $\beta_{LS} \to 0$.

For effectively relating the observables to strong field limit coefficients that carry information about the nature of the black hole, we use the deflection angle formulae and the lens equation to obtain the relationship between the three observations of the relativistic image and strong field limit coefficients. The three observables mentioned above, namely the remaining inner packed images $\theta_\infty$, the separation between the first image and the others $s$, and the ratio of the flux of the first image to the flux of all other images $r_{mag}$, read as \cite{Bozza:2002zj}
\begin{align}
            \theta_{\infty} &= \frac{b_m}{D_{OL}}, \label{Observables1} \\
            s &=\theta_1-\theta_\infty=\theta_\infty^{\frac{\bar{b}-2\pi}{\bar{a}}}, \label{Observables2} \\
            r &= \frac{\mu_{1}}{\sum_{n=2}^{\infty}\mu_{n}}=e^{\frac{2\pi}{\bar{a}}}, \quad r_{\mathrm{mag}}=2.5\log_{10}(r)=\frac{5\pi}{\bar{a}\ln10}. \label{Observables3}
\end{align}

It is worth noting that $\theta_{\infty}$ represents the asymptotic position approached by a set of images in the limit $n \to \infty$ from eq.~(\ref{ri}), i.e., angular radius of the photon sphere.

Therefore, we can predict the above-mentioned observables of the strong lensing for the SRR black hole through strong field limit coefficients and critical impact parameters which carry information of the black hole. In turn, we can also identify the nature of the SRR black hole and lens by the observables.

Another important observable is time delay, whose surprising importance lies in that it is the only dimensional observable \cite{Bozza:2003cp}. When multiple relativistic images are formed, the travel time in different light paths corresponding to different images is generally not the same, but has a time delay, which depends on the nature of the lens. The time delay between the $n^{th}$ and $m^{th}$ images on the same side of the lens could be approximated as \cite{Bozza:2003cp}

\begin{equation}\label{sTdeq}
            \Delta T^s_{p,q} \approx 2\pi(n-m)\frac{\tilde{a} }{\bar{a}}+2\sqrt{\frac{A_m b_m}{B_m}}\left[e^{(\bar{b}-2m\pi\pm\beta_{LS})/2\bar{a}}-e^{(\bar{b}-2n\pi\pm\beta_{LS})/2\bar{a}}\right],
\end{equation}

where
\begin{align}
            \tilde{a} =\frac{\widetilde{R}(0,x_m)}{2\sqrt{\beta_m}}, \quad
            \widetilde{R}(z,x_m)=\frac{2(1-A_0)\sqrt{BA_0}(2C-bD)}{A'\sqrt{C(D^2+4AC)}}\left(1-\frac{1}{\sqrt{A_0}f(z,x_0)}\right).
	\end{align}

Because the value of the first term on the right is much larger than the second term in eq.~(\ref{sTdeq}), we generally consider the dominant order
\begin{equation}
            \Delta\widetilde{T}_{n,m}^{s}=2\pi(n-m)\frac{\tilde{a}}{\bar{a}}.
\end{equation}

The radii of the prograde and retrograde photon spheres are different, so their travel times are obviously different. In this case of different sides, the time delay is \cite{Bozza:2003cp}
\begin{equation}
            \Delta\tilde{T}_{n,m}^o=\frac{\tilde{a}(a)}{\bar{a}(a)}[2\pi n+\beta_{LS}-\bar{b}(a)]+\tilde{b}(a)-\frac{\tilde{a}(-a)}{\bar{a}(-a)}[2\pi m-\beta_{LS}-\bar{b}(-a)]-\tilde{b}(-a).
\end{equation}
where
\begin{align}
            \tilde{b} &= -\pi+\tilde{b}_D\big(r_m\big)+\tilde{b}_R\big(r_m\big)+\tilde{a}\log\bigg(\frac{cr_m^2}{b_m}\bigg), \\
            \tilde{b}_{D} &= 2\tilde{a}\log\frac{2(1-A_{m})}{A_{m}^{\prime}r_{m}}, \\
            \tilde{b}_R(r_m) &= \int_0^1[\widetilde{R}(z , r_m)f(z , r_m)-\widetilde{R}(0 , r_m)f_0(z , r_m)]dz.
\end{align}

We can get the time delay by calculating $\tilde{a}$ and $\tilde{b}$, which are similar to the strong field limit parameters.

Now, we treat M87* and Sgr A* as supermassive SRR black holes to study their observables, and evaluate their differences from the Kerr black hole. In our paper, we take their masses and distances from the Earth as, $M = 6.5\times10^9M_\odot$ and $D_OL = 16.8$ Mpc for M87* \cite{EventHorizonTelescope:2019ggy}, as well as $4 \times10^6M_\odot$ and $D_OL = 8.35$ kpc for Sgr A* \cite{Ghez:2008ms,Gillessen:2008qv,Falcke:2013ola,Reid:2014boa}.
\begin{figure}[t]
\centering
\includegraphics[width=0.45\linewidth]{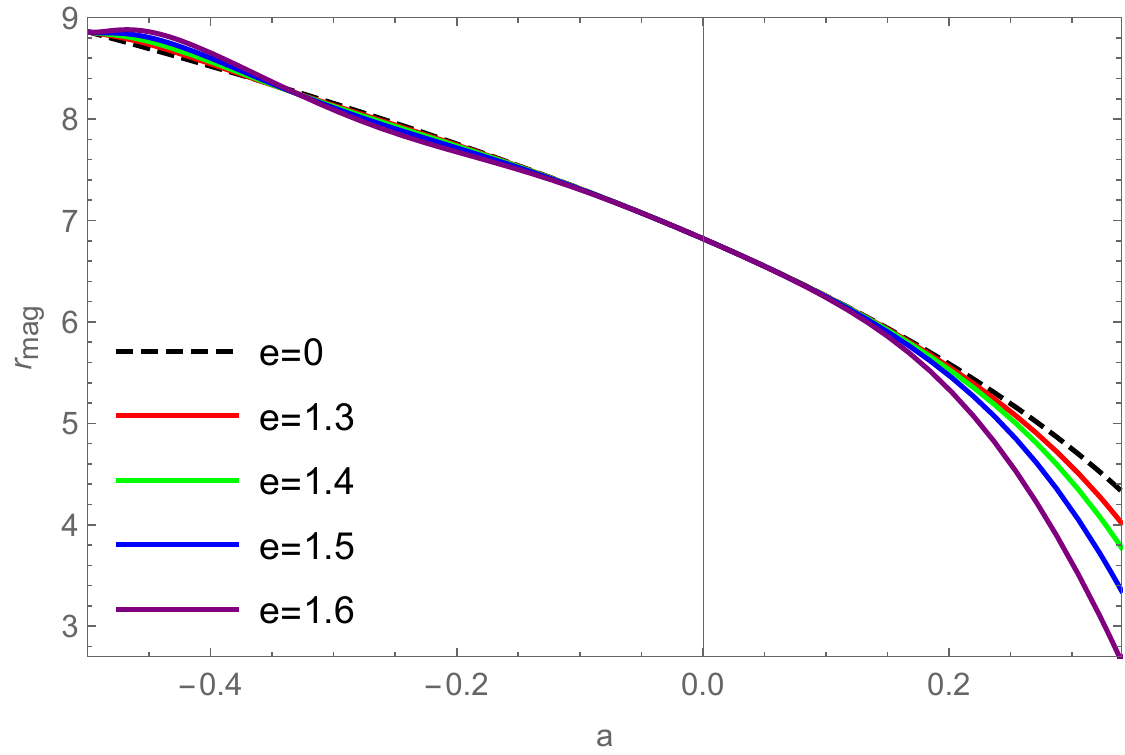}
\caption{The behaviors of lensing observable $r_{mag}$ in strong
field limit.}
\label{sd5}
\end{figure}

\begin{figure}[t]
\centering
\includegraphics[width=0.45\linewidth]{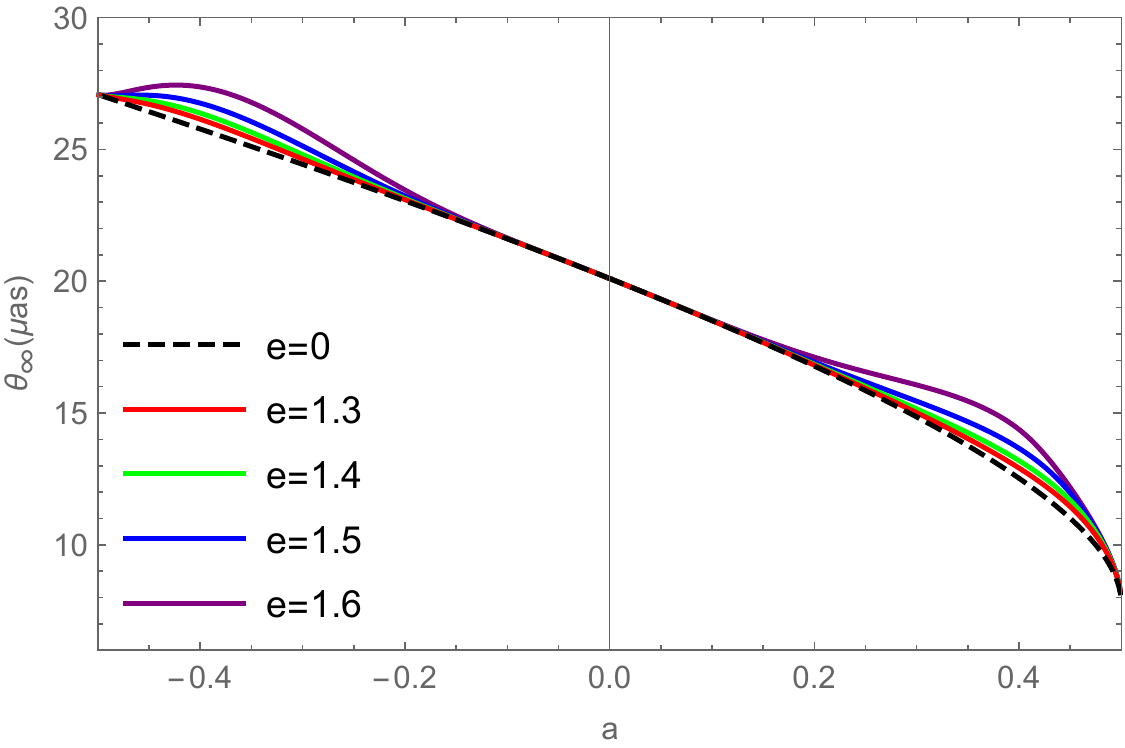}
\includegraphics[width=0.460\linewidth]{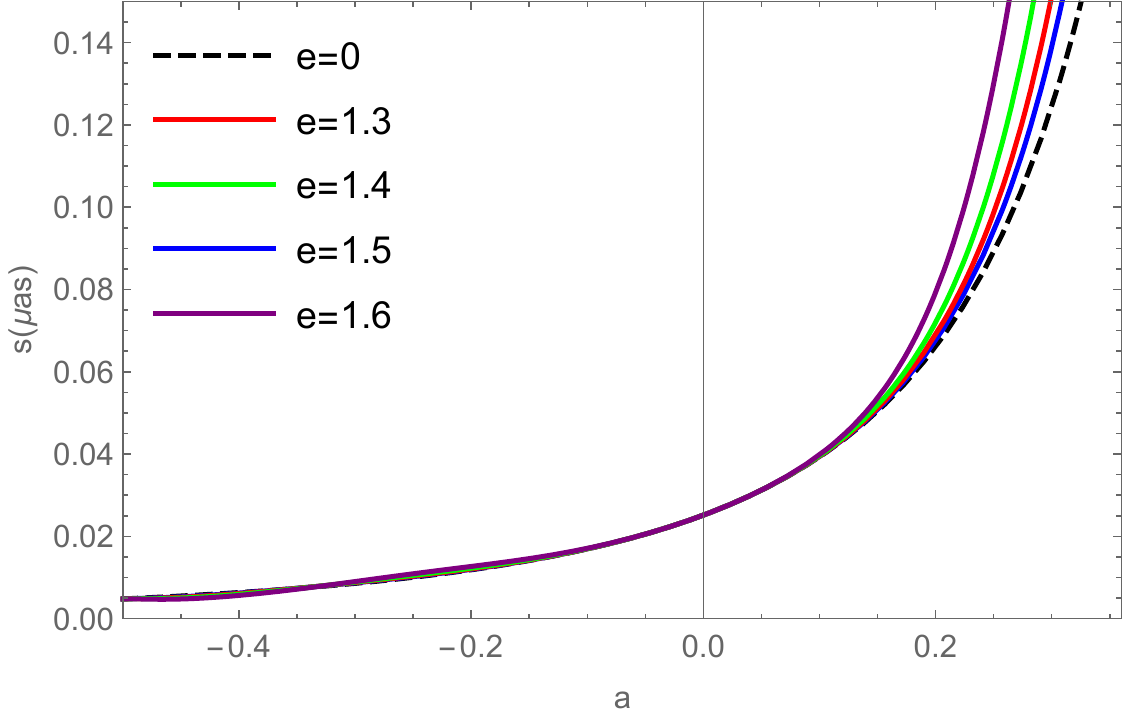}
\caption{The behaviors of lensing observables $\theta_{\infty}$ and $s$ as a function of $a$ by model the M87* as the SRR black hole.}
\label{sd6a7}
\end{figure}

\begin{figure}[t]
\centering
\includegraphics[width=0.45\linewidth]{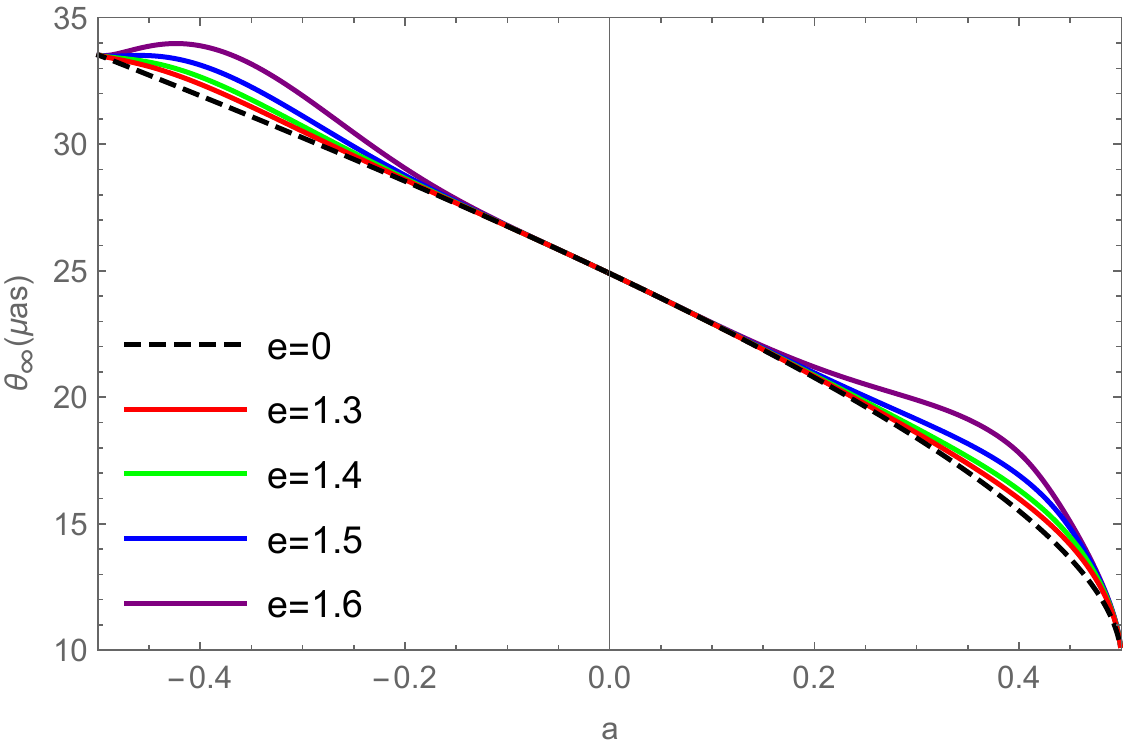}
\includegraphics[width=0.460\linewidth]{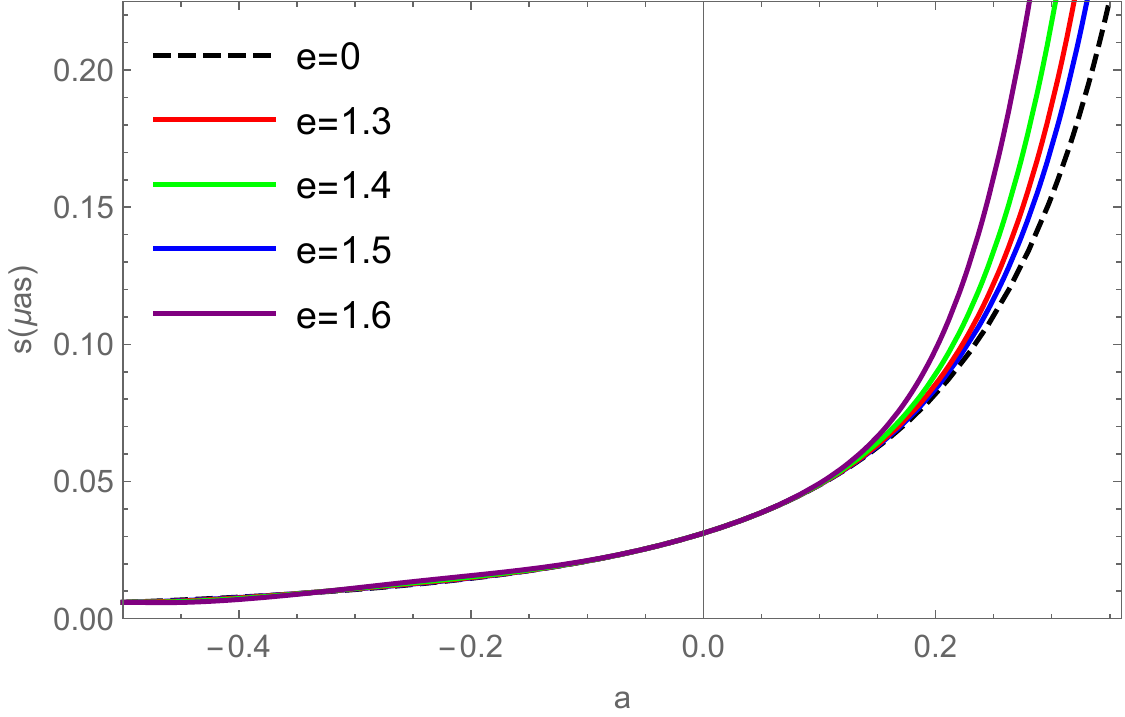}
\caption{The behaviors of lensing observables $\theta_{\infty}$ and $s$ as a function of $a$ by model the Sgr A* as the SRR black hole.}
\label{sd8a9}
\end{figure}

According to eqs.~(\ref{Observables1}), ~(\ref{Observables2}) and~(\ref{Observables3}), it can be seen that the behaviors of the observables have a strong correlation with the strong field limit coefficients $\bar{a}$ and $\bar{b}$, so the characteristics of the strong field limit coefficients will be reflected in the $r_{mag}$, $\theta_{\infty}$ and $s$. The influence of $a$ on $r_{mag}$ and other observables is similar to that of Kerr black holes. As shown in figure~\ref{sd5}, $r_{mag}$ maintains a monotonically decreasing trend for an increase in $a$ when $e$ is small. However, $r_{mag}$ first weakly increases and then decreases for the growth in $a$ when $e$ is large. This characteristic also exists in the change of $\theta$ and $s$ with $a$, i.e., larger $e$ will alter the trend of observables in high spin. In figures~\ref{sd6a7} and~\ref{sd8a9}, we plot the lensing observables $\theta_{\infty}$ and $s$ under the background of M87* and Sgr A*. The left panels of figures~\ref{sd6a7} (M87*) and ~\ref{sd8a9} (Sgr A*) show that $\theta_{\infty}$ has an overall trend of decreasing as the increase of $a$, but $\theta_{\infty}$ first weakly increases and then decreases for the growth in $a$ when $e$ is large. The right panels of figures~\ref{sd6a7} (M87*) and ~\ref{sd8a9} (Sgr A*) show that $s$ first weakly decreases and then increases with the spin $a$. In order to more clearly reflect the influence of $e$ and the difference from Kerr situation, we calculate the deviation of lensing observables between the SRR black hole and Kerr situation.
\begin{figure}[t]
\centering
\includegraphics[width=0.45\linewidth]{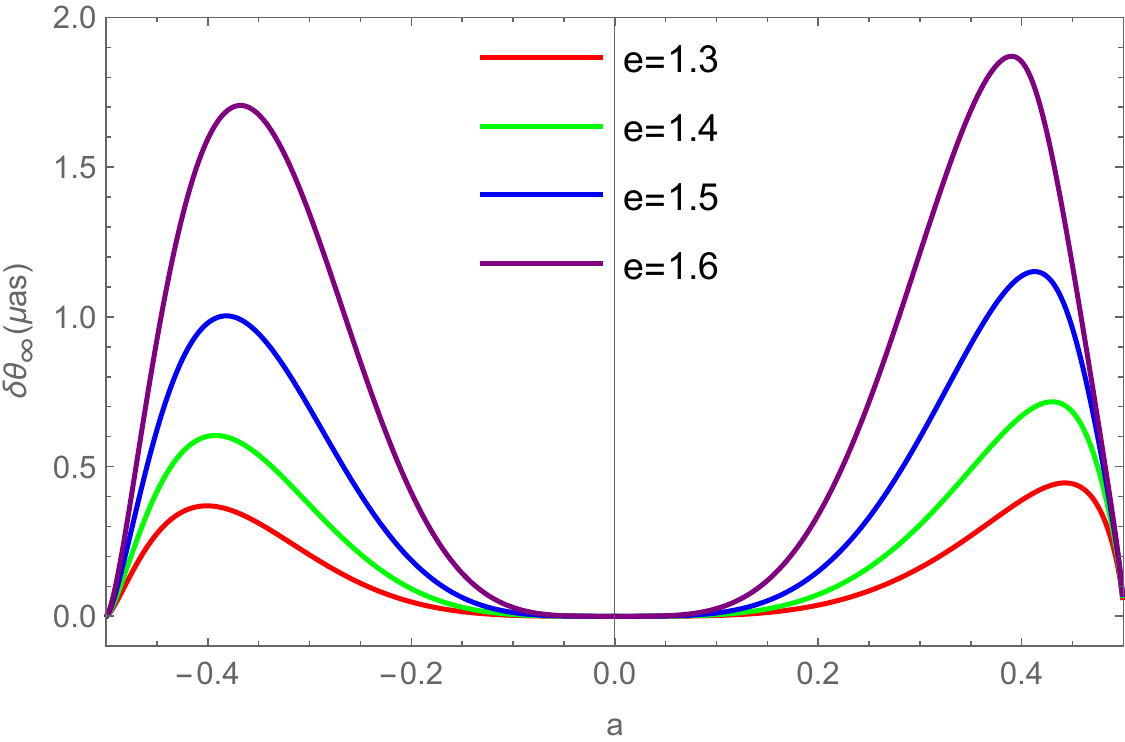}
\includegraphics[width=0.463\linewidth]{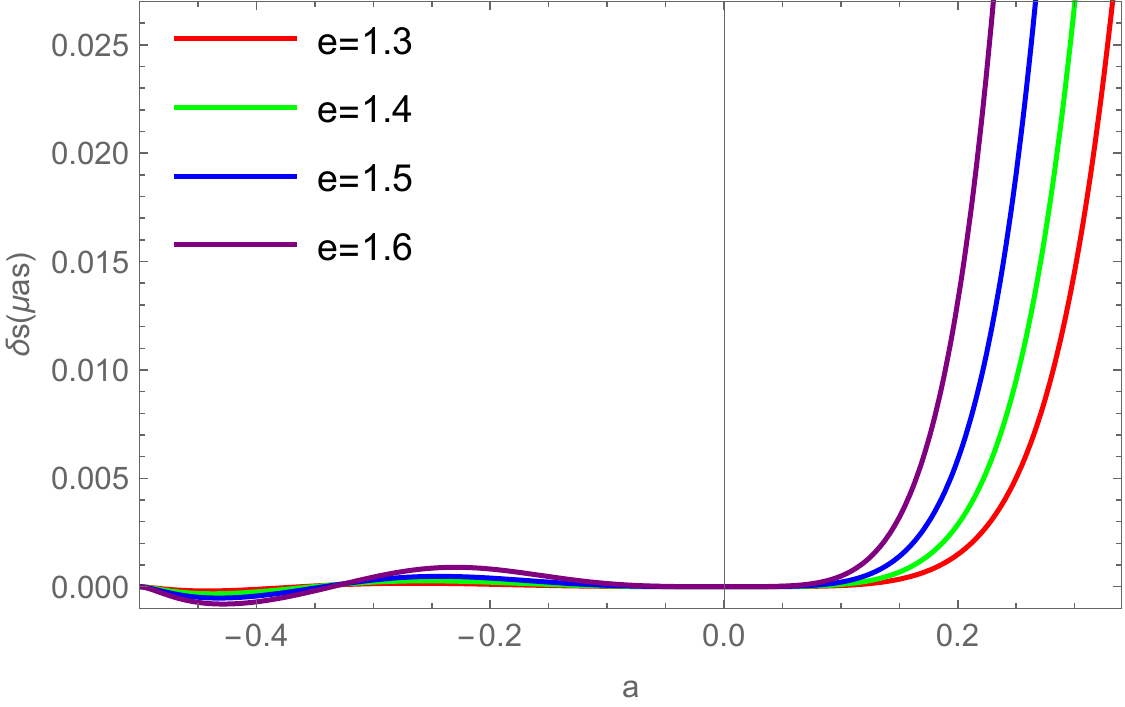}
\caption{The deviations of $\theta_{\infty}$ and $s$ between the SRR black hole and the Kerr black hole in the background of M87*.}
\label{sd10a11}
\end{figure}

\begin{figure}[t]
\centering
\includegraphics[width=0.45\linewidth]{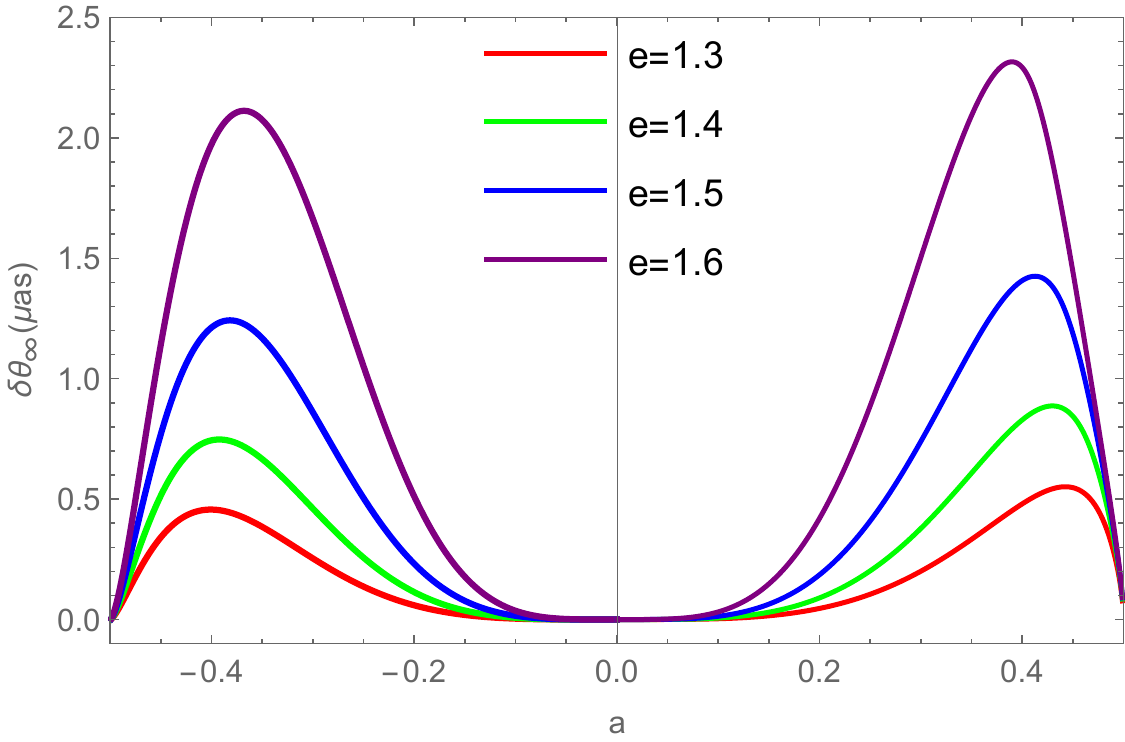}
\includegraphics[width=0.463\linewidth]{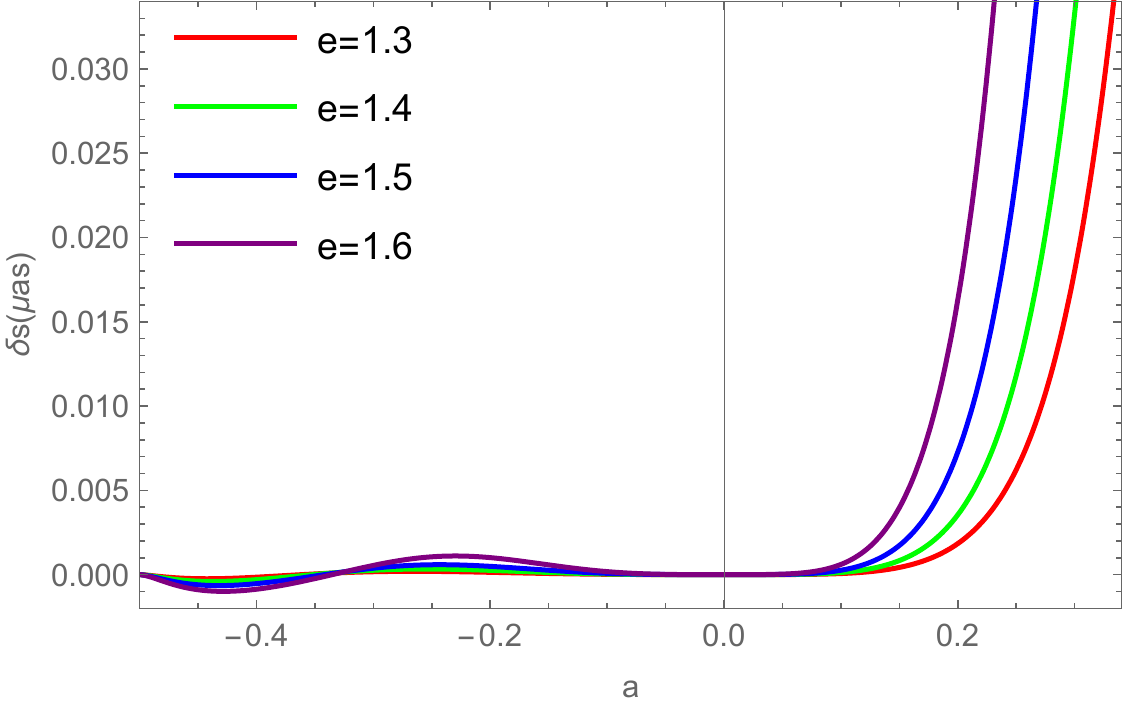}
\caption{The deviations of $\theta_{\infty}$ and $s$ between the SRR black hole and the Kerr black hole in the background of Sgr A*.}
\label{sd12a13}
\end{figure}

We plot the deviation in observables for SRR black holes and Kerr black holes under the background of M87* and Sgr A* in figures~\ref{sd10a11} and~\ref{sd12a13} for the different $e$. For $\theta_{\infty}$, with the increase of $\left | a \right | $, the deviation increases from zero increasing and then decreasing to zero after reaching a peak. And the peak increases with the growth of $e$. With respect to $s$, the spin $a$ and deviation parameter $e$ have tiny impact on retrograde photons, i.e., it causes two weak fluctuations. While for prograde photons, $s$ monotonically increases with the increase of $e$. It can be seen that the presence of $e$ causes $\theta_{\infty}$ to be increased at high spin. For $s$, the presence of $e$ increases $s$ of prograde photons at high spin, while it only has a tiny effect on retrograde photons. This means that we can more easily observe the deviation from the SRR black hole and Kerr situations at high spin.
\begin{center}
\begin{table}[t] \centering
\renewcommand{\arraystretch}{1.2}
\setlength{\tabcolsep}{0.28cm}
\begin{tabular}{|cccccc|}
\hline
\multicolumn{6}{|c|}{$\Delta T_{2,1}(\delta \Delta T_{2,1})$/hrs}               \\
a/M  & e = 0   & e = 1.30       & e = 1.40       & e = 1.50        & e = 1.60        \\ \hline
0.6  & 214.052 & 216.386(2.334) & 218.454(4.402) & 222.656(8.604)  & 231.552(17.500) \\
0.3  & 254.576 & 254.757(0.181) & 254.931(0.355) & 255.320(0.744)  & 256.296(1.720)  \\
0    & 289.758 & 289.758(0)     & 289.758(0)     & 289.758(0)      & 289.758(0)      \\
-0.3 & 321.957 & 322.180(0.224) & 322.393(0.437) & 322.862(0.905)  & 324.028(2.071)  \\
-0.6 & 352.186 & 355.145(2.959) & 357.552(5.366) & 362.202(10.017) & 371.548(19.362) \\ \hline
\end{tabular}
\caption{The time delay of first and second images on the same side for the SRR black hole and its deviation $\delta \Delta T_{2,1}$ from the Kerr situation as the M87*.}
\label{Td1}
\end{table}
\end{center}

\begin{center}
\begin{table}[t] \centering
\renewcommand{\arraystretch}{1.2}
\setlength{\tabcolsep}{0.375cm}
\begin{tabular}{|cccccc|}
\hline
\multicolumn{6}{|c|}{$\Delta T_{2,1}(\delta \Delta T_{2,1})$/min}         \\
a/M  & e = 0  & e = 1.30      & e = 1.40       & e = 1.50      & e = 1.60      \\ \hline
0.6  & 7.903  & 7.990(0.086)  & 8.066(0.163)   & 8.221(0.318)  & 8.550(0.646)  \\
0.3  & 9.400  & 9.406(0.006)  & 9.412(0.012)   & 9.427(0.027)  & 9.463(0.063)  \\
0    & 10.699 & 10.699(0)     & 10.699(0)      & 10.699(0)     & 10.699(0)     \\
-0.3 & 11.888 & 11.896(0.008) & 11.904(0.016)  & 11.921(0.033) & 11.964(0.076) \\
-0.6 & 13.004 & 13.113(0.109) & 13.202(0.0198) & 13.374(0.370) & 13.719(0.715) \\ \hline
\end{tabular}
\caption{The time delay of first and second images on the same side for the SRR black hole and its deviation $\delta \Delta T_{2,1}$ from the Kerr situation as the Sgr A*.}
\label{Td2}
\end{table}
\end{center}

Tables~\ref{Td1} and~\ref{Td2} show the time delay $\Delta T_{2,1}$ between the first and second images on the same side of M87* and Sgr A*, as well as their deviation from Kerr case. We can see that the parameter $a$ has a similar effect on Kerr and SRR black holes, i.e., the time delay of the SRR black hole increases with an increase in $a$. With respect to the time delay of first and second images on the same side, it increases for an increase in the deviation parameter $e$. The time delay  $\Delta T_{1,1}$ of first prograde and first retrograde images on the opposite side, as well as its deviation from the Kerr situation, can be seen in tables~\ref{Td3} and~\ref{Td4}. It shows increase of $\Delta T_{1,1}$ with the increase in $a$, but to our surprise, $\Delta T_{1,1}$ decreases as $e$ increases. The deviations are similar to other lensing observables, both $\delta \Delta T_{2,1}$ and $\delta \Delta T_{1,1}$ increase when we increase $e$ as can be seen from tables~\ref{Td1},~\ref{Td2},~\ref{Td3} and~\ref{Td4}.
\begin{center}
\begin{table}[t] \centering
\renewcommand{\arraystretch}{1.2}
\setlength{\tabcolsep}{0.25cm}
\begin{tabular}{|cccccc|}
\hline
\multicolumn{6}{|c|}{$\Delta T_{1,1}(\delta \Delta T_{1,1})$/hrs}                       \\
a/M  & e = 0    & e = 1.30       & e = 1.40         & e = 1.50        & e = 1.60        \\ \hline
0.6  & 120.6047 & 112.072(-8.53) & 103.857(-16.748) & 85.784(-34.820) & 41.882(-78.722) \\
0.3  & 61.49555 & 61.335(-0.164) & 61.167(-0.329)   & 60.792(-0.704)  & 59.812(-1.684)  \\
0    & 0        & 0              & 0                & 0               & 0               \\
-0.3 & -61.4956 & -61.331(0.165) & -61.167(0.329)   & -60.792(0.704)  & -59.812(1.684)  \\
-0.6 & -120.605 & -112.072(8.53) & -103.857(16.748) & -85.784(34.820) & -41.882(78.722) \\ \hline
\end{tabular}
\caption{The time delay of first prograde and first retrograde images on the opposite side for the SRR black hole and its deviation $\delta \Delta T_{1,1}$ from the Kerr situation as the M87*.}
\label{Td3}
\end{table}
\end{center}

\begin{center}
\begin{table}[t] \centering
\renewcommand{\arraystretch}{1.2}
\setlength{\tabcolsep}{0.415cm}
\begin{tabular}{|cccccc|}
\hline
\multicolumn{6}{|c|}{$\Delta T_{1,1}(\delta \Delta T_{1,1})$/min}             \\
a/M  & e = 0  & e = 1.30      & e = 1.40      & e = 1.50      & e = 1.60      \\ \hline
0.6  & 4.453  & 4.138(-0.315) & 3.835(-0.618) & 3.167(-1.286) & 1.546(-2.907) \\
0.3  & 2.271  & 2.265(-0.006) & 2.258(-0.012) & 2.245(-0.026) & 2.208(-0.062) \\
0    & 0      & 0             & 0             & 0             & 0             \\
-0.3 & -2.271 & -2.265(0.006) & -2.258(0.012) & -2.245(0.026) & -2.208(0.062) \\
-0.6 & -4.453 & -4.138(0.315) & -3.835(0.618) & -3.167(1.286) & -1.546(2.907) \\ \hline
\end{tabular}
\caption{The time delay of first prograde and first retrograde images on the opposite side for the SRR black hole and its deviation $\delta \Delta T_{1,1}$ from the Kerr situation as the Sgr A*.}
\label{Td4}
\end{table}
\end{center}

In brief, the deviation parameter $e$ has a significant effect on the observables, which is closely related to the spin parameter $a$. The influence is negligible when $a$ is tiny, while it is often obvious if $a$ takes larger value.

\section{Weak gravitational lensing}
\label{Weak gravitational lensing}
The GBT is a fundamental formula which links the curvature of a surface to its underlying topology. Firstly, Gibbon and Werner applied the GBT to calculate the deflection angle of light at the weak field limit for spherically symmetric black hole spacetime \cite{Gibbons:2008rj}. Later, Werner \cite{Werner:2012rc} and Ono \cite{Ono:2017pie} extended it to Kerr spacetime in Kerr-Randers optical geometry and spatial metric, respectively. Their methods have been continuously promoted and are widely used in various black hole spacetimes \cite{Ishihara:2016vdc,Crisnejo:2018uyn,Ovgun:2018fnk,Ovgun:2018tua,Ono:2019hkw,Ovgun:2019wej,Javed:2019rrg,Javed:2019ynm,Javed:2019kon,Zhu:2019ura,Kumar:2019pjp,Li:2020ozr,Pantig:2021zqe,Uniyal:2022vdu,Pantig:2022toh,Kumaran:2022soh,Parbin:2023zik}. According to Ono's method, we compute the deflection of light along the equatorial plane ($\theta = \pi / 2$) of the SRR black hole spacetime.

First, let us consider the coordinate system centered on the black hole (L). Assuming that the distance between the observer (O) and the source (S) is infinite, the deflection angle can be expressed as \cite{Ishihara:2016vdc,Ishihara:2016sfv,Ono:2017pie}
\begin{equation}\label{wlda1}
            \alpha_{Wd} =\Psi_O - \Psi_S + \Phi_{OS},
\end{equation}
where $\Phi_{OS}$ is the angular separation between the observer and the source. $\Psi_S$ and $\Psi_O$ are angular coordinates that are measured at the receiver position and the source position, respectively.

\begin{figure}[t]
\centering
\includegraphics[width=0.6\linewidth]{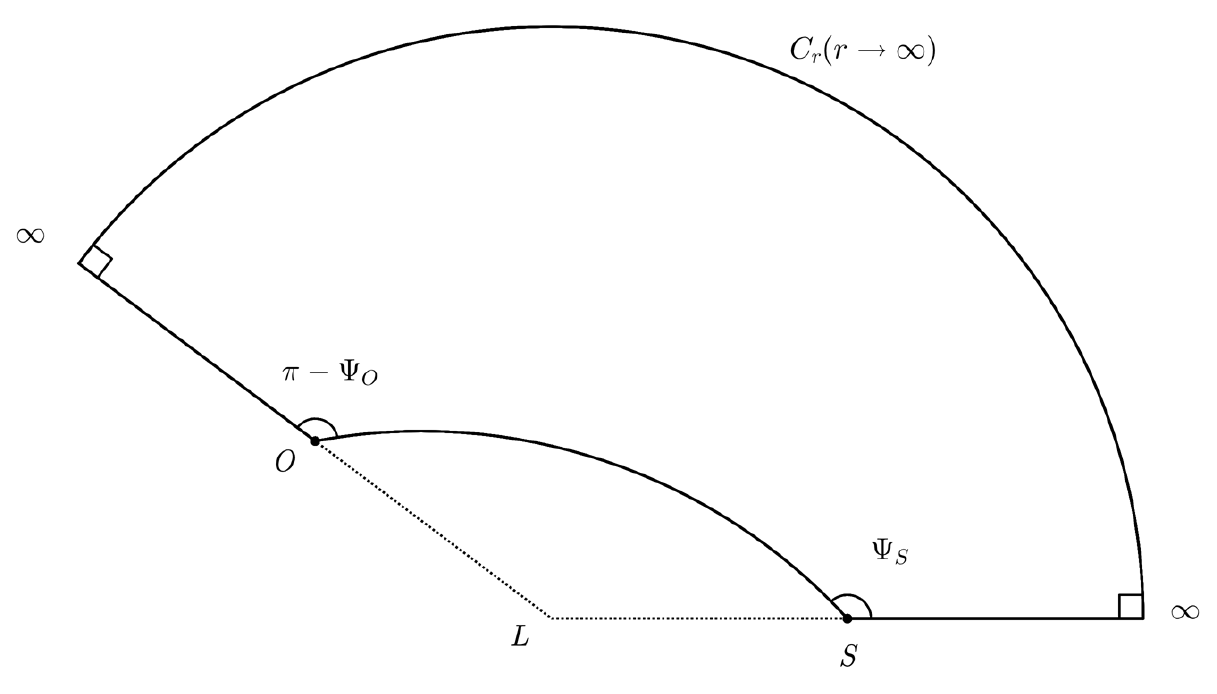}
\caption{Quadrilateral ${_{O}^{\infty} \Box _{S}^{\infty}}$ embedded in the curved space.}
\label{GBT}
\end{figure}

As shown in figure~\ref{GBT}, the quadrilateral $_{O}^{\infty} \Box _{S}^{\infty}$ which is embedded in a 3-dimensional Riemannian manifold $_{}^{(3)} \mathcal{M}$, consists of a space curve of light rays from $S$ to $O$, two outgoing radial lines from $O$ and from $S$, as well as a circular arc segment $C_r$ of coordinate radius $r_C$ ($r_C \to \infty$) \cite{Ishihara:2016vdc}. Applying GBT in the above scenario, the light deflection angle~(\ref{wlda1}) can be written as \cite{Ono:2017pie}
\begin{equation}\label{wlda2}
            \alpha_{Wd} =-\iint_{_{O}^{\infty} \Box _{S}^{\infty}}^{} \mathcal{K} dS \,+\int_{S}^{O} k_g dl,
\end{equation}
where $\mathcal{K}$ and $k_g$ respectively imply the Gaussian curvature of the light propagation surface and the geodesic curvature of the light curve. $d S $ is the infinitesimal area element of the surface, and $dl$ is the infinitesimal line element of the arc. It should be noted that $dl > 0$ represents the prograde motion of photons and $dl < 0$ represents the retrograde motion.

For obtaining the deflection angle of light near the considered black hole, we first need to calculate the Gaussian curvature $\mathcal{K}$ of the propagating light. These light rays can be described as space curves which are depicted by spatial metric on the 3-dimensional Riemannian manifold $_{}^{(3)} \mathcal{M}$ \cite{Ono:2017pie}. By solving ${d s}^2=0$, we get
\begin{equation}
           dt = \pm \sqrt{\gamma _{ij} {dx}^i {dx}^j} + N_i {dx}^i.
\end{equation}

Here, we will use the same form of line elements as eq.~(\ref{simmetric}), so $\gamma_{ij}dx^{i}dx^{j}$ and $N_idx^i$ read as
\begin{align}
           \gamma_{ij}dx^{i}dx^{j} &= \frac{B(r)}{A(r)}dr^{2} + \frac{4A(r)C(r)+D^{2}(r)}{4A^{2}(r)}d\phi^{2}, \\
           N_idx^i &= -\frac{D(r)}{2A(r)}d\phi.
\end{align}

With this spatial metric, the Gaussian curvature of propagating light is defined as \cite{Werner:2012rc,Ono:2017pie}
\begin{equation}
           \mathcal{K} = \frac{_{}^{(3)} R_{r\phi r\phi }}{\gamma } = \frac{1}{\sqrt{\gamma } }  \left [\frac{\partial  }{\partial \phi } \left ( \frac{\sqrt{\gamma } }{\gamma_{rr}}   {_{}^{(3)} \Gamma  _{r r}^{\phi}} \right ) - \frac{\partial  }{\partial r } \left ( \frac{\sqrt{\gamma } }{\gamma_{rr}}   {_{}^{(3)} \Gamma  _{r \phi}^{\phi}}\right ) \right ],
\end{equation}
where $\gamma = det(\gamma_{ij})$. For the SRR black hole, in the weak field limit and slowly rotating approximation, $\mathcal{K}$ is computed as follows:
\begin{equation}
		  \begin{split}
           \mathcal{K} &= -\frac {1024 M}{(2 - e)^9 r^3} + \frac {4608 e M}{(2 - e)^9 r^3}
 - \frac{9216 e^2 M}{(2 - e)^9 r^3}  + \frac {1536 M^2}{(2 - e)^9 r^4} \\
&- \frac {7912 e M^2}{(2 - e)^9 r^4} + \frac {13824 e^2 M^2}{(2 - e)^9 r^4}
 + \mathcal{O}(e^3, a^2, M^3, \frac{1}{r^{7}}).
            \end{split}
\end{equation}

The integral of Gaussian curvature over the closed quadrilateral ${_{O}^{\infty} \Box _{S}^{\infty}}$ reads \cite{Ono:2017pie}
\begin{equation}\label{Gc1}
           \iint_{_{O}^{\infty} \Box _{S}^{\infty}}^{} \mathcal{K} dS = \int_{\phi_S}^{\phi_O} \int_{ \infty }^{r_0} \mathcal{K} \sqrt{\gamma}  drd\phi,
\end{equation}
where $r_0$ is the closed distance to the black hole. In weak field limit and the slow rotation approximation, the equation of light orbits can be approximated as $u = \frac{\sin \phi }{b } + \mathcal{O}(M)$. Here we introduce $u = 1 / r $ and the impact parameter $b$. Hence, we can rewrite eq.~(\ref{Gc1}) as
\begin{equation}
           \iint_{_{O}^{\infty} \Box _{S}^{\infty}}^{} \mathcal{K} dS = \int_{\phi_S}^{\phi_O} \int_{ 0 }^{ \frac{\sin \phi }{b }} - \frac{ \mathcal{K} \sqrt{\gamma} }{u^2} du d\phi,
\end{equation}
which for the metric~(\ref{metric}) reads
\begin{equation}
		  \begin{split}
                     \iint \mathcal{K} dS &=
                     \sqrt{1-b^2 {u_O}^2}\left [ \frac{1024 M}{b (2-e)^9} - \frac{4608 M  e}{b (2-e)^9} + \frac{9216 M e^2}{b (2-e)^9}\right ] \\
                     & + \sqrt{1-b^2 {u_S}^2}\left [ \frac{1024 M}{b (2-e)^9} - \frac{4608 M  e}{b (2-e)^9} + \frac{9216 M e^2}{b (2-e)^9}\right ] \\
                     & + \sqrt{b^2 {u_O}^2-b^4 {u_O}^4} \left [ \frac{384 M^2}{b^2 (2-e)^9} - \frac{1478 e M^2}{b^2 (2-e)^9} + \frac{3456 e^2 M^2}{b^2 (2-e)^9} \right ]\\
                     & + \sqrt{b^2 {u_S}^2-b^4 {u_S}^4} \left [ \frac{384 M^2}{b^2 (2-e)^9} - \frac{1478 e M^2}{b^2 (2-e)^9} + \frac{3456 e^2 M^2}{b^2 (2-e)^9} \right ]\\
                     & + \left [\pi-\arccos(b u_{O})-\arccos(b u_{S})\right ] \left [\frac{384 M^2}{b^2 (2-e)^9} - \frac{1478 e M^2}{b^2 (2-e)^9} + \frac{3456 e^2 M^2}{b^2 (2-e)^9}\right ] \\
                     & + \mathcal{O}(e^3, a^2, M^3, \frac{1}{r^{7}}),
		  \end{split}
\end{equation}
where $u_O$ and $u_S$ are the reciprocal of the observer and source distances from the black hole, respectively. We also use the approximation $\cos {\phi_O} = - \sqrt{1 - b^2 {u_O}^2}$, $\cos {\phi_S} = \sqrt{1 - b^2 {u_S}^2}$. Next, the geodesic curvature in the manifold $_{}^{(3)} \mathcal{M}$ can be expressed as \cite{Ono:2017pie}
\begin{equation}
                    k_g = - \frac{1} {\sqrt{\gamma \gamma^{\theta \theta}} } \beta_{\phi , r},
\end{equation}
which for the black hole metric~(\ref{metric}) yields
\begin{equation}
		  \begin{split}
             k_g &=  - \frac{1024 a M}{(2 - e)^9 r^3}
                        + \frac { 4608 a e M}{(2 - e)^9 r^3} - \frac {9216 a e^2 M}{(2 - e)^9 r^3}- \frac{ 1024 a M^2}{(2 - e)^9 r^4} \\
                        & + \frac {4608 a e M^2}{(2 - e)^9 r^4}
                        - \frac { 9216 a e^2 M^2}{(2 - e)^9 r^4}
                        +\mathcal{O}(e^3, a^3, M^3, \frac{1}{r^{7}}).
 		  \end{split}
\end{equation}

The contribution of the geodesic curvature exists as the eq.~(\ref{wlda2}), and we use a linear approximation of the photon orbit as $1/r = u = \sin {\phi} / b + \mathcal{O}(M)$. The path integral of geodesic curvature reads
\begin{equation}
		  \begin{split}
            \int_{S}^{O} k_g dl &= - \sqrt{1-b^2 {u_O}^2}\left [  \frac{512 a M ( 18e^2 - 9 e +2 ) }{b^2 (2-e^9)}\right ] \\
            & - \sqrt{1-b^2 {u_S}^2} \left [\frac{512 a M  (18e^2 - 9 e +2) }{b^2 (2-e^9)}\right ] \\
            & - \sqrt{b^2 {u_O}^2-b^4 {u_O}^4} \left [\frac{256 a M^2  (18 e^2-9 e+2)}{b^3 (2-e^9)}\right ] \\
            & - \sqrt{b^2 {u_S}^2-b^4 {u_S}^4} \left [\frac{256 a M^2  (18 e^2-9 e+2)}{b^3 (2-e^9)}\right ]\\
            & - \left [\pi-\arccos(b u_{O})-\arccos(b u_{S})\right ] \left [\frac{256 a M^2  (18 e^2-9 e+2)}{b^3 (2-e^9)}\right ] \\
            & + \mathcal{O}(e^3, a^3, M^3, \frac{1}{r^{7}}).
		  \end{split}
\end{equation}

In the limits of far distant observer and source, i.e., $u_O \to 0$ and $u_S \to 0$, the deflection angle for the SRR black hole in the weak field limits reads as follows:
\begin{equation} \label{wdf}
	\begin{split}
		\alpha_{Wd} & = \frac{2048 M}{b (2-e)^9} - \frac{9216 e M}{b (2-e)^9} + \frac{18432 e^2 M}{b (2-e)^9} + \frac{384 \pi  M^2}{b^2 (2-e)^9}\\
		& - \frac{1478 \pi e M^2}{b^2 (2-e)^9} + \frac{3456 \pi e^2 M^2}{b^2 (2-e)^9} - \frac{2048 a M}{b^2 (2-e)^9} + \frac{9216 a e M}{b^2 (2-e)^9}\\
		& - \frac{512 \pi a M^2}{b^3 (2-e)^9} + \frac{2304 \pi  a e M^2}{b^3 (2-e)^9} - \frac{18432 a e^2 M}{b^2 (2-e)^9} - \frac{4608 \pi  a e^2 M^2}{b^3 (2-e)^9}.
	\end{split}
\end{equation}

\begin{figure}[t]
\centering
\includegraphics[width=0.45\linewidth]{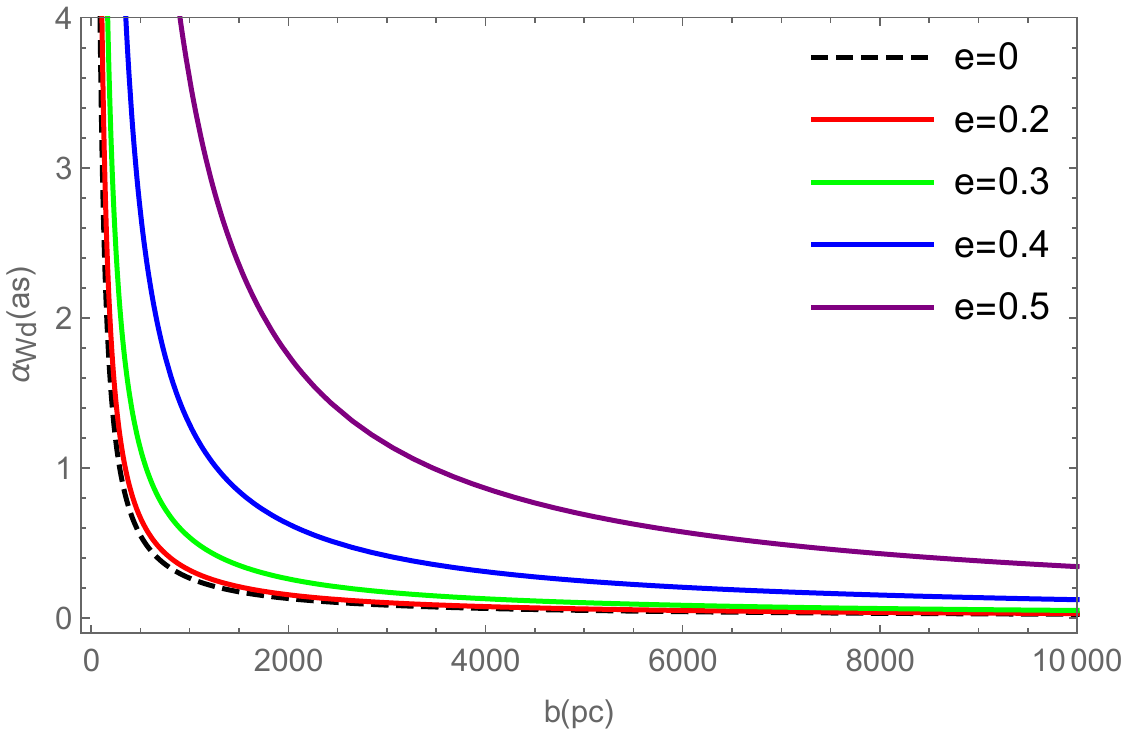}
\includegraphics[width=0.487\linewidth]{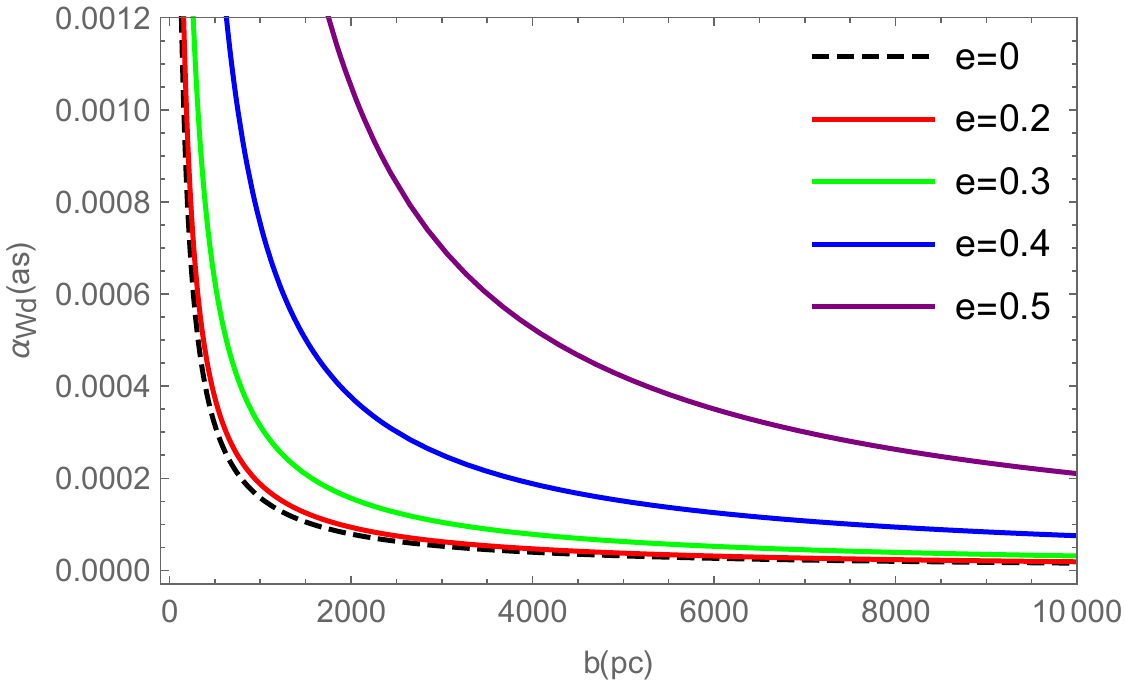}
\caption{The behaviors of weak deflection angle $\alpha_{Wd}$ as a function of the impact parameter $b$ for the M87* (left panel) and  Sgr A* (right panel) with different values of $e$.}
\label{wd1a2}
\end{figure}

\begin{figure}[t]
\centering
\includegraphics[width=0.47\linewidth]{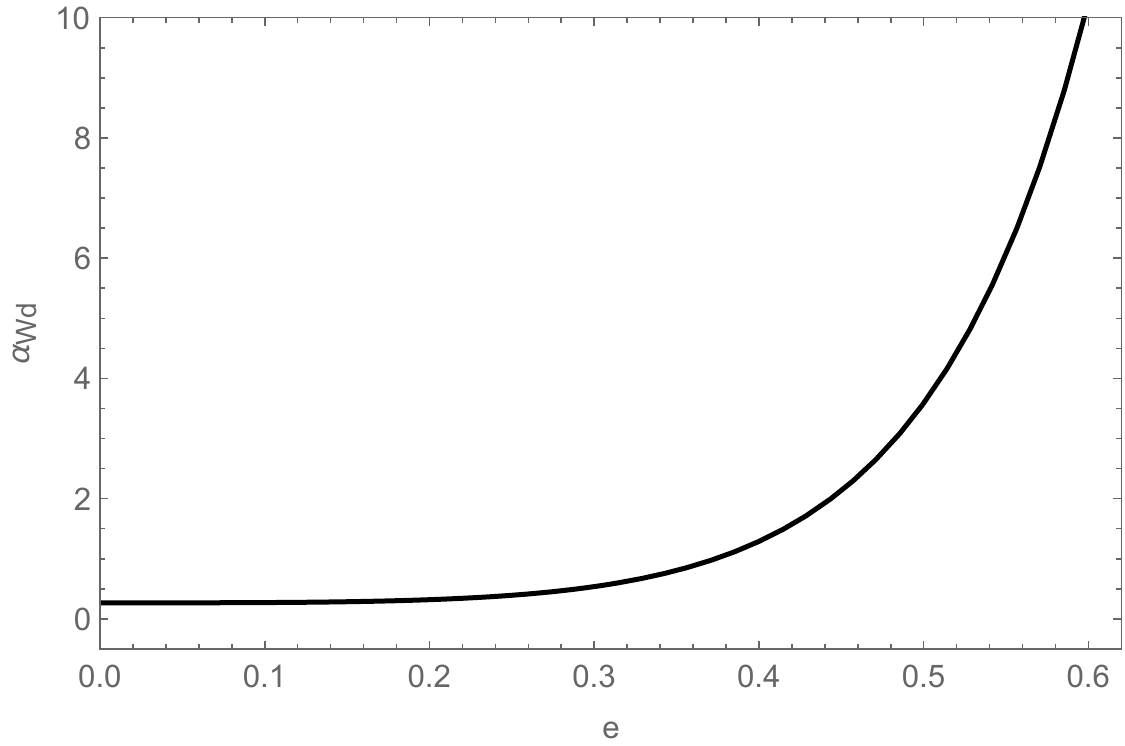}
\caption{The behaviors of weak deflection angle $\alpha_{Wd}$ as a function of the deviation parameter $e$ for the SRR black hole.}
\label{wd3}
\end{figure}

Here, we use M87* and Sgr A* as black hole models and calculate the influence of different parameters $e$ on the deflection angle under the weak field limits. Figure~\ref{wd1a2} shows the variation of the weak deflection angle $\alpha_{Wd}$ with the impact parameter $b$ with different deviation parameter $e$ for M87* and Sgr A*, respectively. Compared to the Kerr case, the existence of the deviation parameter $e$ increases the weak deflection angle $\alpha_{Wd}$ of light ray near the black hole. Furthermore, we plot the change of $\alpha_{Wd}$ when we increase $e$ in figure~\ref{wd3}. As we expected in the strong deflection, the deflection angle of light around the black hole increases for the growth of $e$. In addition, with increasing deviation parameter $e$, eq.~(\ref{wdf}) and figure~\ref{wd3} imply that $\alpha_{Wd}$ presents a trend of first slowly increasing and then rapidly increasing.

\section{Conclusion} 
\label{Conclusion}
In this paper, we have studied the gravitational lensing effect of the SRR black hole. First, we have verified that the parameter $l$ of the conformal factor does not affect the gravitational lensing effect. Then, in the strong and weak field limit, we have calculated the radius of the photon sphere, deflection angle of light and lensing observables which are described by the mass ($M$), spin parameter ($a$), and deviation parameter ($e$). In the strong field limit, the influence of $a$ on the gravitational lensing effect of the SRR black hole is similar to that of the Kerr black hole. As $a$ increases, both the radius of the photon sphere and the light deflection angle $\alpha_{Sd}(\theta)$ decrease when the impact parameter $b$ is fixed. Moreover, $\alpha_{Sd}(\theta)$ increases monotonically for an increase in the deviation parameter $e$.

To investigate the SRR black hole astrophysical relevance, we regard M87* and Sgr A* as models of lens. After that, we have computed the observables which include the position of asymptotic relativistic images $\theta_{\infty}$, separation $s$, magnification $r_{mag}$, and time delays between different images. Overall, the observables of the SRR black hole are similar to those in the Kerr case, except when the parameters $a$ and $e$ are larger. Furthermore, we have evaluated the deviation of observables of the SRR black hole from Kerr case under the M87* and Sgr A* backgrounds. With the growth of $\left | a \right |$, $\delta \theta_{\infty}$ first increases up to a peak and then decreases. The value of $\delta s$ shows a descending trend, and then the value ascends, before descending again, when we increase $a$. As $a$ increases further, $\delta s$ shows a monotonous increasing trend. The increase of $e$ will increase the deviations on the lensing observables. For the observables and the deviations mentioned above, we have found that Sgr A* is easier to observe than M87* under the same conditions.

Later, we have explored the time delays for Sgr A* and M87*, as well as their deviations ($\delta \Delta T_{2,1}$ and $\delta \Delta T_{1,1}$) from the SRR black hole and Kerr case. The most intuitive thing is the increase of $\Delta T_{2,1}$ and the decrease of $\Delta T_{1,1}$ for an increase in $e$. Both the deviations of time delay $\delta \Delta T_{2,1}$ and $\delta \Delta T_{1,1}$ increase with the increase in $e$. We have found that the time delay for M87* and its deviation from the Kerr black hole are much longer than those for Sgr A*. These results imply that M87* is a better choice for testing or distinguishing different black holes. For both the same side and the opposite side time delays, although the images on the opposite side are more pronounced, the time delay on the same side is greater. Therefore, M87* is a better choice when the observation precision of the time delay is not high enough.

Finally, we have used GBT to calculate the light deflection angle in the weak field limit. It shows that the spin $a$ has an extremely small impact on the deflection angle, but the deflection angle of light increases slowly at first and then increases dramatically for an increase in deviation parameter $e$.

 In the future, assuming that the SRR black hole is selected as one of the candidates for M87* and Sgr A*, we can make distinguishable predictions from Kerr black holes under both strong and weak gravity conditions. In a period when general relativity is not powerful enough and a complete quantum gravity theory has not yet emerged, the SRR black hole solves the theoretical flaws (central singularity, mass inflation of the Cauchy horizon and CTCs) of the Kerr black hole under general relativity from a phenomenological perspective. Our work provides predictions of the observables of the SRR black hole, which may contribute to distinguishing SRR black holes and Kerr black holes in astronomical observations.

\acknowledgments

This work was supported by Yunnan Fundamental Research Projects (Grant No. 202301AS070029), and Yunnan High-level Talent Training Support Plan Young \& Elite Talents Project (Grant No. YNWR-QNBJ-2018-360).




\begin{thebibliography}{99}

\bibitem{EventHorizonTelescope:2019dse}
K.~Akiyama \textit{et al.} [Event Horizon Telescope],
\emph{First M87 Event Horizon Telescope Results. I. The Shadow of the Supermassive Black Hole},
\emph{Astrophys. J. Lett.} \textbf{875} (2019), L1
[arXiv:1906.11238 [astro-ph.GA]].

\bibitem{EventHorizonTelescope:2019ggy}
K.~Akiyama \textit{et al.} [Event Horizon Telescope],
\emph{First M87 Event Horizon Telescope Results. VI. The Shadow and Mass of the Central Black Hole},
\emph{Astrophys. J. Lett.} \textbf{875} (2019) no.1, L6
[arXiv:1906.11243 [astro-ph.GA]].

\bibitem{EventHorizonTelescope:2019pgp}
K.~Akiyama \textit{et al.} [Event Horizon Telescope],
\emph{First M87 Event Horizon Telescope Results. V. Physical Origin of the Asymmetric Ring},
\emph{Astrophys. J. Lett.} \textbf{875} (2019) no.1, L5
[arXiv:1906.11242 [astro-ph.GA]].

\bibitem{EventHorizonTelescope:2022wkp}
K.~Akiyama \textit{et al.} [Event Horizon Telescope],
\emph{First Sagittarius A* Event Horizon Telescope Results. I. The Shadow of the Supermassive Black Hole in the Center of the Milky Way},
\emph{Astrophys. J. Lett.} \textbf{930} (2022) no.2, L12
[arXiv:2311.08680 [astro-ph.HE]].

\bibitem{EventHorizonTelescope:2022xqj}
K.~Akiyama \textit{et al.} [Event Horizon Telescope],
\emph{First Sagittarius A* Event Horizon Telescope Results. VI. Testing the Black Hole Metric},
\emph{Astrophys. J. Lett.} \textbf{930} (2022) no.2, L17
[arXiv:2311.09484 [astro-ph.HE]].

\bibitem{Cui:2023uyb}
Y.~Cui, K.~Hada, T.~Kawashima, M.~Kino, W.~Lin, Y.~Mizuno, H.~Ro, M.~Honma, K.~Yi and J.~Yu, \textit{et al.}
\emph{Precessing jet nozzle connecting to a spinning black hole in M87},
\emph{Nature} \textbf{621} (2023), 711-715
[arXiv:2310.09015 [astro-ph.HE]].


\bibitem{Penrose:1964wq}
R.~Penrose,
\emph{Gravitational collapse and space-time singularities},
\emph{Phys. Rev. Lett.} \textbf{14} (1965), 57-59

\bibitem{Hawking:1973uf}
S.~W.~Hawking and G.~F.~R.~Ellis,
\emph{The Large Scale Structure of Space-Time},
Cambridge University Press, Cambridge (1973).

\bibitem{Wald:1984rg}
R.~M.~Wald,
\emph{General Relativity},
Chicago Univ. Pr., Chicago (1984).




\bibitem{Poisson:1989zz}
E.~Poisson and W.~Israel,
\emph{Inner-horizon instability and mass inflation in black holes},
\emph{Phys. Rev. Lett.} \textbf{63} (1989), 1663-1666

\bibitem{Franzin:2022wai}
E.~Franzin, S.~Liberati, J.~Mazza and V.~Vellucci,
\emph{Stable rotating regular black holes},
\emph{Phys. Rev. D} \textbf{106} (2022) no.10, 104060
[arXiv:2207.08864 [gr-qc]].

\bibitem{Englert:1976ep}
F.~Englert, C.~Truffin and R.~Gastmans,
\emph{Conformal Invariance in Quantum Gravity},
\emph{Nucl. Phys. B} \textbf{117} (1976), 407-432

\bibitem{Narlikar:1977nf}
J.~V.~Narlikar and A.~K.~Kembhavi,
\emph{Space-Time Singularities and Conformal Gravity},
\emph{Lett. Nuovo Cim.} \textbf{19} (1977), 517-520

\bibitem{Bars:2013yba}
I.~Bars, P.~Steinhardt and N.~Turok,
\emph{Local Conformal Symmetry in Physics and Cosmology},
\emph{Phys. Rev. D} \textbf{89} (2014) no.4, 043515
[arXiv:1307.1848 [hep-th]].

\bibitem{DominisPrester:2013rpi}
P.~Dominis Prester,
\emph{Curing black hole singularities with local scale invariance},
\emph{Adv. Math. Phys.} \textbf{2016} (2016), 6095236
[arXiv:1309.1188 [hep-th]].

\bibitem{Bambi:2016yne}
C.~Bambi, L.~Modesto, S.~Porey and L.~Rachwa\l{},
\emph{Black hole evaporation in conformal gravity},
\emph{JCAP} \textbf{09} (2017), 033
[arXiv:1611.05582 [gr-qc]].

\bibitem{Bartelmann:1999yn}
M.~Bartelmann and P.~Schneider,
\emph{Weak gravitational lensing},
\emph{Phys. Rept.} \textbf{340} (2001), 291-472
[arXiv:astro-ph/9912508 [astro-ph]].

\bibitem{Perlick:2004tq}
V.~Perlick,
\emph{Gravitational lensing from a spacetime perspective},
\emph{Living Rev. Rel.} \textbf{7} (2004), 9

\bibitem{Bozza:2010xqn}
V.~Bozza,
\emph{Gravitational Lensing by Black Holes},
\emph{Gen. Rel. Grav.} \textbf{42} (2010), 2269-2300
[arXiv:0911.2187 [gr-qc]].


\bibitem{Aubourg:1993wb}
E.~Aubourg, P.~Bareyre, S.~Brehin, M.~Gros, M.~Lachieze-Rey, B.~Laurent, E.~Lesquoy, C.~Magneville, A.~Milsztain and L.~Moscoso, \textit{et al.}
\emph{Evidence for gravitational microlensing by dark objects in the galactic halo},
\emph{Nature} \textbf{365} (1993), 623-625






\bibitem{Clowe:2006eq}
D.~Clowe, M.~Bradac, A.~H.~Gonzalez, M.~Markevitch, S.~W.~Randall, C.~Jones and D.~Zaritsky,
\emph{A direct empirical proof of the existence of dark matter},
\emph{Astrophys. J. Lett.} \textbf{648} (2006), L109-L113
[arXiv:astro-ph/0608407 [astro-ph]].




\bibitem{Eiroa:2002mk}
E.~F.~Eiroa, G.~E.~Romero and D.~F.~Torres,
\emph{Reissner-Nordstrom black hole lensing},
\emph{Phys. Rev. D} \textbf{66} (2002), 024010
[arXiv:gr-qc/0203049 [gr-qc]].

\bibitem{Gyulchev:2006zg}
G.~N.~Gyulchev and S.~S.~Yazadjiev,
\emph{Kerr-Sen dilaton-axion black hole lensing in the strong deflection limit},
\emph{Phys. Rev. D} \textbf{75} (2007), 023006
[arXiv:gr-qc/0611110 [gr-qc]].

\bibitem{Bin-Nun:2009hct}
A.~Y.~Bin-Nun,
\emph{Relativistic Images in Randall-Sundrum II Braneworld Lensing},
\emph{Phys. Rev. D} \textbf{81} (2010), 123011
[arXiv:0912.2081 [gr-qc]].

\bibitem{Kraniotis:2010gx}
G.~V.~Kraniotis,
\emph{Precise analytic treatment of Kerr and Kerr-(anti) de Sitter black holes as gravitational lenses},
\emph{Class. Quant. Grav.} \textbf{28} (2011), 085021
[arXiv:1009.5189 [gr-qc]].

\bibitem{Wei:2011nj}
S.~W.~Wei, Y.~X.~Liu, C.~E.~Fu and K.~Yang,
\emph{Strong field limit analysis of gravitational lensing in Kerr-Taub-NUT spacetime},
\emph{JCAP}, \textbf{10} (2012), 053
[arXiv:1104.0776 [hep-th]].

\bibitem{Kraniotis:2014paa}
G.~V.~Kraniotis,
\emph{Gravitational lensing and frame dragging of light in the Kerr-Newman and the Kerr-Newman-(anti) de Sitter black hole spacetimes},
\emph{Gen. Rel. Grav.} \textbf{46} (2014) no.11, 1818
[arXiv:1401.7118 [gr-qc]].

\bibitem{Zhao:2016kft}
S.~S.~Zhao and Y.~Xie,
\emph{Strong field gravitational lensing by a charged Galileon black hole},
\emph{JCAP} \textbf{07} (2016), 007
[arXiv:1603.00637 [gr-qc]].

\bibitem{Chakraborty:2016lxo}
S.~Chakraborty and S.~SenGupta,
\emph{Strong gravitational lensing --- A probe for extra dimensions and Kalb-Ramond field},
\emph{JCAP} \textbf{07} (2017), 045
[arXiv:1611.06936 [gr-qc]].

\bibitem{Zhao:2017cwk}
S.~S.~Zhao and Y.~Xie,
\emph{Strong deflection gravitational lensing by a modified Hayward black hole},
\emph{Eur. Phys. J. C} \textbf{77} (2017) no.5, 272
[arXiv:1704.02434 [gr-qc]].

\bibitem{Jusufi:2018jof}
K.~Jusufi, A.~\"Ovg\"un, J.~Saavedra, Y.~V\'asquez and P.~A.~Gonz\'alez,
\emph{Deflection of light by rotating regular black holes using the Gauss-Bonnet theorem},
\emph{Phys. Rev. D} \textbf{97} (2018) no.12, 124024
[arXiv:1804.00643 [gr-qc]].

\bibitem{Wang:2019cuf}
C.~Y.~Wang, Y.~F.~Shen and Y.~Xie,
\emph{Weak and strong deflection gravitational lensings by a charged Horndeski black hole},
\emph{JCAP} \textbf{04} (2019), 022
[arXiv:1902.03789 [gr-qc]].

\bibitem{Jusufi:2019caq}
K.~Jusufi, M.~Jamil, H.~Chakrabarty, Q.~Wu, C.~Bambi and A.~Wang,
\emph{Rotating regular black holes in conformal massive gravity},
\emph{Phys. Rev. D} \textbf{101} (2020) no.4, 044035
[arXiv:1911.07520 [gr-qc]].

\bibitem{Kumar:2020sag}
R.~Kumar, S.~U.~Islam and S.~G.~Ghosh,
\emph{Gravitational lensing by charged black hole in regularized $4D$ Einstein\textendash{}Gauss\textendash{}Bonnet gravity},
\emph{Eur. Phys. J. C} \textbf{80} (2020) no.12, 1128
[arXiv:2004.12970 [gr-qc]].

\bibitem{Ghosh:2020spb}
S.~G.~Ghosh, R.~Kumar and S.~U.~Islam,
\emph{Parameters estimation and strong gravitational lensing of nonsingular Kerr-Sen black holes},
JCAP \textbf{03} (2021), 056
[arXiv:2011.08023 [gr-qc]].






\bibitem{Bhadra:2003zs}
A.~Bhadra,
\emph{Gravitational lensing by a charged black hole of string theory},
\emph{Phys. Rev. D} \textbf{67} (2003), 103009
[arXiv:gr-qc/0306016 [gr-qc]].

\bibitem{Horvath:2011xr}
Z.~Horvath, L.~A.~Gergely, Z.~Keresztes, T.~Harko and F.~S.~N.~Lobo,
\emph{Constraining Ho\v{r}ava-Lifshitz gravity by weak and strong gravitational lensing},
\emph{Phys. Rev. D} \textbf{84} (2011), 083006
[arXiv:1105.0765 [gr-qc]].

\bibitem{Sahu:2015dea}
S.~Sahu, K.~Lochan and D.~Narasimha,
\emph{Gravitational lensing by self-dual black holes in loop quantum gravity},
\emph{Phys. Rev. D} \textbf{91} (2015), 063001
[arXiv:1502.05619 [gr-qc]].

\bibitem{Badia:2017art}
J.~Bad\'\i{}a and E.~F.~Eiroa,
\emph{Gravitational lensing by a Horndeski black hole},
\emph{Eur. Phys. J. C} \textbf{77} (2017) no.11, 779
[arXiv:1707.02970 [gr-qc]].

\bibitem{Allahyari:2019jqz}
A.~Allahyari, M.~Khodadi, S.~Vagnozzi and D.~F.~Mota,
\emph{Magnetically charged black holes from non-linear electrodynamics and the Event Horizon Telescope},
\emph{JCAP} \textbf{02} (2020), 003
[arXiv:1912.08231 [gr-qc]].

\bibitem{Li:2020zxi}
Z.~Li and T.~Zhou,
\emph{Kerr black hole surrounded by a cloud of strings and its weak gravitational lensing in Rastall gravity},
\emph{Phys. Rev. D} \textbf{104} (2021) no.10, 104044
[arXiv:2001.01642 [gr-qc]].

\bibitem{Afrin:2021imp}
M.~Afrin, R.~Kumar and S.~G.~Ghosh,
\emph{Parameter estimation of hairy Kerr black holes from its shadow and constraints from M87*},
\emph{Mon. Not. Roy. Astron. Soc.} \textbf{504} (2021), 5927-5940
[arXiv:2103.11417 [gr-qc]].

\bibitem{Afrin:2021wlj}
M.~Afrin and S.~G.~Ghosh,
\emph{Testing Horndeski Gravity from EHT Observational Results for Rotating Black Holes},
\emph{Astrophys. J.} \textbf{932} (2022) no.1, 51
[arXiv:2110.05258 [gr-qc]].

\bibitem{Kuang:2022ojj}
X.~M.~Kuang, Z.~Y.~Tang, B.~Wang and A.~Wang,
\emph{Constraining a modified gravity theory in strong gravitational lensing and black hole shadow observations},
\emph{Phys. Rev. D} \textbf{106} (2022) no.6, 064012
[arXiv:2206.05878 [gr-qc]].

\bibitem{Kuang:2022xjp}
X.~M.~Kuang and A.~\"Ovg\"un,
\emph{Strong gravitational lensing and shadow constraint from M87* of slowly rotating Kerr-like black hole},
\emph{Annals Phys.} \textbf{447} (2022), 169147
[arXiv:2205.11003 [gr-qc]].

\bibitem{Vagnozzi:2022moj}
S.~Vagnozzi, R.~Roy, Y.~D.~Tsai, L.~Visinelli, M.~Afrin, A.~Allahyari, P.~Bambhaniya, D.~Dey, S.~G.~Ghosh and P.~S.~Joshi, \textit{et al.}
\emph{Horizon-scale tests of gravity theories and fundamental physics from the Event Horizon Telescope image of Sagittarius A},
\emph{Class. Quant. Grav.} \textbf{40} (2023) no.16, 165007
[arXiv:2205.07787 [gr-qc]].

\bibitem{Pantig:2022ely}
R.~C.~Pantig and A.~\"Ovg\"un,
\emph{Testing dynamical torsion effects on the charged black hole\textquoteright{}s shadow, deflection angle and greybody with M87* and Sgr. A* from EHT},
\emph{Annals Phys.} \textbf{448} (2023), 169197
[arXiv:2206.02161 [gr-qc]].

\bibitem{Soares:2023err}
A.~R.~Soares, R.~L.~L.~Vit\'oria and C.~F.~S.~Pereira,
\emph{Gravitational lensing in a topologically charged Eddington-inspired Born\textendash{}Infeld spacetime},
\emph{Eur. Phys. J. C} \textbf{83} (2023) no.10, 903
[arXiv:2305.11105 [gr-qc]].




\bibitem{Falcke:1999pj}
H.~Falcke, F.~Melia and E.~Agol,
\emph{Viewing the shadow of the black hole at the galactic center},
\emph{Astrophys. J. Lett.} \textbf{528} (2000), L13
[arXiv:astro-ph/9912263 [astro-ph]].

\bibitem{Takahashi:2004xh}
R.~Takahashi,
\emph{Shapes and positions of black hole shadows in accretion disks and spin parameters of black holes},
\emph{J. Korean Phys. Soc.} \textbf{45} (2004), S1808-S1812
[arXiv:astro-ph/0405099 [astro-ph]].

\bibitem{Bambi:2008jg}
C.~Bambi and K.~Freese,
\emph{Apparent shape of super-spinning black holes},
\emph{Phys. Rev. D} \textbf{79} (2009), 043002
[arXiv:0812.1328 [astro-ph]].

\bibitem{Hioki:2009na}
K.~Hioki and K.~I.~Maeda,
\emph{Measurement of the Kerr Spin Parameter by Observation of a Compact Object's Shadow},
\emph{Phys. Rev. D} \textbf{80} (2009), 024042
[arXiv:0904.3575 [astro-ph.HE]].

\bibitem{Amarilla:2011fx}
L.~Amarilla and E.~F.~Eiroa,
\emph{Shadow of a rotating braneworld black hole},
\emph{Phys. Rev. D} \textbf{85} (2012), 064019
[arXiv:1112.6349 [gr-qc]].

\bibitem{Grenzebach:2014fha}
A.~Grenzebach, V.~Perlick and C.~L\"ammerzahl,
\emph{Photon Regions and Shadows of Kerr-Newman-NUT Black Holes with a Cosmological Constant},
\emph{Phys. Rev. D} \textbf{89} (2014) no.12, 124004
[arXiv:1403.5234 [gr-qc]].

\bibitem{Cunha:2015yba}
P.~V.~P.~Cunha, C.~A.~R.~Herdeiro, E.~Radu and H.~F.~Runarsson,
\emph{Shadows of Kerr black holes with scalar hair},
\emph{Phys. Rev. Lett.} \textbf{115} (2015) no.21, 211102
[arXiv:1509.00021 [gr-qc]].

\bibitem{Abdujabbarov:2015pqp}
A.~Abdujabbarov, B.~Toshmatov, Z.~Stuchl\'\i{}k and B.~Ahmedov,
\emph{Shadow of the rotating black hole with quintessential energy in the presence of plasma},
\emph{Int. J. Mod. Phys. D} \textbf{26} (2016) no.06, 1750051
[arXiv:1512.05206 [gr-qc]].

\bibitem{Abdujabbarov:2016hnw}
A.~Abdujabbarov, M.~Amir, B.~Ahmedov and S.~G.~Ghosh,
\emph{Shadow of rotating regular black holes},
\emph{Phys. Rev. D} \textbf{93} (2016) no.10, 104004
[arXiv:1604.03809 [gr-qc]].

\bibitem{Younsi:2016azx}
Z.~Younsi, A.~Zhidenko, L.~Rezzolla, R.~Konoplya and Y.~Mizuno,
\emph{New method for shadow calculations: Application to parametrized axisymmetric black holes},
\emph{Phys. Rev. D} \textbf{94} (2016) no.8, 084025
[arXiv:1607.05767 [gr-qc]].

\bibitem{Tsukamoto:2017fxq}
N.~Tsukamoto,
\emph{Black hole shadow in an asymptotically-flat, stationary, and axisymmetric spacetime: The Kerr-Newman and rotating regular black holes},
\emph{Phys. Rev. D} \textbf{97} (2018) no.6, 064021
[arXiv:1708.07427 [gr-qc]].

\bibitem{Cunha:2018acu}
P.~V.~P.~Cunha and C.~A.~R.~Herdeiro,
\emph{Shadows and strong gravitational lensing: a brief review},
\emph{Gen. Rel. Grav.} \textbf{50} (2018) no.4, 42
[arXiv:1801.00860 [gr-qc]].

\bibitem{Shaikh:2018lcc}
R.~Shaikh, P.~Kocherlakota, R.~Narayan and P.~S.~Joshi,
\emph{Shadows of spherically symmetric black holes and naked singularities},
\emph{Mon. Not. Roy. Astron. Soc.} \textbf{482} (2019) no.1, 52-64
[arXiv:1802.08060 [astro-ph.HE]].

\bibitem{Shaikh:2018kfv}
R.~Shaikh,
\emph{Shadows of rotating wormholes},
\emph{Phys. Rev. D} \textbf{98} (2018) no.2, 024044
[arXiv:1803.11422 [gr-qc]].

\bibitem{Gralla:2019xty}
S.~E.~Gralla, D.~E.~Holz and R.~M.~Wald,
\emph{Black Hole Shadows, Photon Rings, and Lensing Rings},
\emph{Phys. Rev. D} \textbf{100} (2019) no.2, 024018
[arXiv:1906.00873 [astro-ph.HE]].

\bibitem{Vagnozzi:2019apd}
S.~Vagnozzi and L.~Visinelli,
\emph{Hunting for extra dimensions in the shadow of M87*},
\emph{Phys. Rev. D} \textbf{100} (2019) no.2, 024020
[arXiv:1905.12421 [gr-qc]].

\bibitem{Konoplya:2019fpy}
R.~A.~Konoplya, T.~Pappas and A.~Zhidenko,
\emph{Einstein-scalar\textendash{}Gauss-Bonnet black holes: Analytical approximation for the metric and applications to calculations of shadows},
\emph{Phys. Rev. D} \textbf{101} (2020) no.4, 044054
[arXiv:1907.10112 [gr-qc]].

\bibitem{Konoplya:2021slg}
R.~A.~Konoplya and A.~Zhidenko,
\emph{Shadows of parametrized axially symmetric black holes allowing for separation of variables},
\emph{Phys. Rev. D} \textbf{103} (2021) no.10, 104033
[arXiv:2103.03855 [gr-qc]].

\bibitem{Ghosh:2022gka}
R.~Ghosh, M.~Rahman and A.~K.~Mishra,
\emph{Regularized stable Kerr black hole: cosmic censorships, shadow and quasi-normal modes},
\emph{Eur. Phys. J. C} \textbf{83} (2023) no.1, 91
[arXiv:2209.12291 [gr-qc]].






\bibitem{Darwin:1959wai}
C. Darwin, \emph{The gravity field of a particle}, \emph{Proc. R. Soc. Lond. A.}, 
\textbf{249}, (1959) 180.

\bibitem{Virbhadra:1999nm}
K.~S.~Virbhadra and G.~F.~R.~Ellis,
\emph{Schwarzschild black hole lensing},
\emph{Phys. Rev. D} \textbf{62} (2000), 084003
[arXiv:astro-ph/9904193 [astro-ph]].

\bibitem{Virbhadra:2002ju}
K.~S.~Virbhadra and G.~F.~R.~Ellis,
\emph{Gravitational lensing by naked singularities},
\emph{Phys. Rev. D} \textbf{65} (2002), 103004

\bibitem{Virbhadra:2008ws}
K.~S.~Virbhadra,
\emph{Relativistic images of Schwarzschild black hole lensing},
\emph{Phys. Rev. D} \textbf{79} (2009), 083004
[arXiv:0810.2109 [gr-qc]].

\bibitem{Virbhadra:2022iiy}
K.~S.~Virbhadra,
\emph{Distortions of images of Schwarzschild lensing},
\emph{Phys. Rev. D} \textbf{106} (2022) no.6, 064038
[arXiv:2204.01879 [gr-qc]].

\bibitem{Virbhadra:2022ybp}
K.~S.~Virbhadra,
\emph{Compactness of supermassive dark objects at galactic centers},
[arXiv:2204.01792 [gr-qc]].

\bibitem{Bozza:2002zj}
V.~Bozza,
\emph{Gravitational lensing in the strong field limit},
\emph{Phys. Rev. D} \textbf{66} (2002), 103001
[arXiv:gr-qc/0208075 [gr-qc]].

\bibitem{Bozza:2002af}
V.~Bozza,
\emph{Quasiequatorial gravitational lensing by spinning black holes in the strong field limit},
\emph{Phys. Rev. D} \textbf{67} (2003), 103006
[arXiv:gr-qc/0210109 [gr-qc]].

\bibitem{Bozza:2003cp}
V.~Bozza and L.~Mancini,
\emph{Time delay in black hole gravitational lensing as a distance estimator},
\emph{Gen. Rel. Grav.} \textbf{36} (2004), 435-450
[arXiv:gr-qc/0305007 [gr-qc]].



\bibitem{Whisker:2004gq}
R.~Whisker,
\emph{Strong gravitational lensing by braneworld black holes},
\emph{Phys. Rev. D} \textbf{71} (2005), 064004
[arXiv:astro-ph/0411786 [astro-ph]].

\bibitem{Chen:2009eu}
S.~Chen and J.~Jing,
\emph{Strong field gravitational lensing in the deformed Ho\v{r}ava-Lifshitz black hole},
\emph{Phys. Rev. D} \textbf{80} (2009), 024036
[arXiv:0905.2055 [gr-qc]].

\bibitem{Liu:2010wh}
Y.~Liu, S.~Chen and J.~Jing,
\emph{Strong gravitational lensing in a squashed Kaluza-Klein black hole spacetime},
\emph{Phys. Rev. D} \textbf{81} (2010), 124017
[arXiv:1003.1429 [gr-qc]].

\bibitem{Ding:2010dc}
C.~Ding, S.~Kang, C.~Y.~Chen, S.~Chen and J.~Jing,
\emph{Strong gravitational lensing in a noncommutative black-hole spacetime},
\emph{Phys. Rev. D} \textbf{83} (2011), 084005
[arXiv:1012.1670 [gr-qc]].

\bibitem{Sotani:2015ewa}
H.~Sotani and U.~Miyamoto,
\emph{Strong gravitational lensing by an electrically charged black hole in Eddington-inspired Born-Infeld gravity},
\emph{Phys. Rev. D} \textbf{92} (2015) no.4, 044052
[arXiv:1508.03119 [gr-qc]].

\bibitem{Tsukamoto:2016qro}
N.~Tsukamoto,
\emph{Strong deflection limit analysis and gravitational lensing of an Ellis wormhole},
\emph{Phys. Rev. D} \textbf{94} (2016) no.12, 124001
[arXiv:1607.07022 [gr-qc]].

\bibitem{Hsieh:2021scb}
T.~Hsieh, D.~S.~Lee and C.~Y.~Lin,
\emph{Strong gravitational lensing by Kerr and Kerr-Newman black holes},
\emph{Phys. Rev. D} \textbf{103} (2021) no.10, 104063
[arXiv:2101.09008 [gr-qc]].

\bibitem{Ghosh:2022mka}
S.~Ghosh and A.~Bhattacharyya,
\emph{Analytical study of gravitational lensing in Kerr-Newman black-bounce spacetime},
\emph{JCAP} \textbf{11} (2022), 006
[arXiv:2206.09954 [gr-qc]].

\bibitem{Tsukamoto:2022tmm}
N.~Tsukamoto,
\emph{Affine perturbation series of the deflection angle of a ray near the photon sphere of a Reissner-Nordstr\"om black hole},
\emph{Phys. Rev. D} \textbf{106} (2022) no.8, 084025
[arXiv:2208.10197 [gr-qc]].

\bibitem{Tsukamoto:2022uoz}
N.~Tsukamoto,
\emph{Gravitational lensing by using the 0th order of affine perturbation series of the deflection angle of a ray near a photon sphere},
\emph{Eur. Phys. J. C} \textbf{83} (2023) no.4, 284
[arXiv:2211.04239 [gr-qc]].

\bibitem{AbhishekChowdhuri:2023ekr}
A.~Chowdhuri, S.~Ghosh and A.~Bhattacharyya,
\emph{A review on analytical studies in Gravitational Lensing},
\emph{Front. Phys.} \textbf{11} (2023), 1113909
[arXiv:2303.02069 [gr-qc]].

\bibitem{Soares:2023uup}
A.~R.~Soares, C.~F.~S.~Pereira, R.~L.~L.~Vit\'oria and E.~M.~Rocha,
\emph{Holonomy corrected Schwarzschild black hole lensing},
\emph{Phys. Rev. D} \textbf{108} (2023) no.12, 124024
[arXiv:2309.05106 [gr-qc]].





\bibitem{Eiroa:2013nra}
E.~F.~Eiroa and C.~M.~Sendra,
\emph{Regular phantom black hole gravitational lensing},
\emph{Phys. Rev. D} \textbf{88} (2013) no.10, 103007
[arXiv:1308.5959 [gr-qc]].

\bibitem{Lu:2016gsf}
X.~Lu, F.~W.~Yang and Y.~Xie,
\emph{Strong gravitational field time delay for photons coupled to Weyl tensor in a Schwarzschild black hole},
\emph{Eur. Phys. J. C} \textbf{76} (2016) no.7, 357
[arXiv:1606.02932 [gr-qc]].

\bibitem{Cavalcanti:2016mbe}
R.~T.~Cavalcanti, A.~G.~da Silva and R.~da Rocha,
\emph{Strong deflection limit lensing effects in the minimal geometric deformation and Casadio\textendash{}Fabbri\textendash{}Mazzacurati solutions},
\emph{Class. Quant. Grav.} \textbf{33} (2016) no.21, 215007
[arXiv:1605.01271 [gr-qc]].

\bibitem{Islam:2021ful}
S.~U.~Islam, J.~Kumar and S.~G.~Ghosh,
\emph{Strong gravitational lensing byrotating Simpson-Visser black holes},
\emph{JCAP} \textbf{10} (2021), 013
[arXiv:2104.00696 [gr-qc]].

\bibitem{Hsieh:2021rru}
T.~Hsieh, D.~S.~Lee and C.~Y.~Lin,
\emph{Gravitational time delay effects by Kerr and Kerr-Newman black holes in strong field limits},
\emph{Phys. Rev. D} \textbf{104} (2021) no.10, 104013
[arXiv:2108.05006 [gr-qc]].





\bibitem{Gibbons:2008rj}
G.~W.~Gibbons and M.~C.~Werner,
\emph{Applications of the Gauss-Bonnet theorem to gravitational lensing},
\emph{Class. Quant. Grav.} \textbf{25} (2008), 235009
[arXiv:0807.0854 [gr-qc]].

\bibitem{Werner:2012rc}
M.~C.~Werner,
\emph{Gravitational lensing in the Kerr-Randers optical geometry},
\emph{Gen. Rel. Grav.} \textbf{44} (2012), 3047-3057
[arXiv:1205.3876 [gr-qc]].

\bibitem{Ono:2017pie}
T.~Ono, A.~Ishihara and H.~Asada,
\emph{Gravitomagnetic bending angle of light with finite-distance corrections in stationary axisymmetric spacetimes},
\emph{Phys. Rev. D} \textbf{96} (2017) no.10, 104037
[arXiv:1704.05615 [gr-qc]].


\bibitem{Bambi:2016wdn}
C.~Bambi, L.~Modesto and L.~Rachwa\l{},
\emph{Spacetime completeness of non-singular black holes in conformal gravity},
\emph{JCAP} \textbf{05} (2017), 003
[arXiv:1611.00865 [gr-qc]].




\bibitem{Tsukamoto:2016jzh}
N.~Tsukamoto,
\emph{Deflection angle in the strong deflection limit in a general asymptotically flat, static, spherically symmetric spacetime},
\emph{Phys. Rev. D} \textbf{95} (2017) no.6, 064035
[arXiv:1612.08251 [gr-qc]].


\bibitem{Bozza:2001xd}
V.~Bozza, S.~Capozziello, G.~Iovane and G.~Scarpetta,
\emph{Strong field limit of black hole gravitational lensing},
\emph{Gen. Rel. Grav.} \textbf{33} (2001), 1535-1548
[arXiv:gr-qc/0102068 [gr-qc]].

\bibitem{Bozza:2008ev}
V.~Bozza,
\emph{A Comparison of approximate gravitational lens equations and a proposal for an improved new one},
\emph{Phys. Rev. D} \textbf{78} (2008), 103005
[arXiv:0807.3872 [gr-qc]].



\bibitem{Ghez:2008ms}
A.~M.~Ghez, S.~Salim, N.~N.~Weinberg, J.~R.~Lu, T.~Do, J.~K.~Dunn, K.~Matthews, M.~Morris, S.~Yelda and E.~E.~Becklin, \textit{et al.}
\emph{Measuring Distance and Properties of the Milky Way's Central Supermassive Black Hole with Stellar Orbits},
\emph{Astrophys. J.} \textbf{689} (2008), 1044-1062
[arXiv:0808.2870 [astro-ph]].

\bibitem{Gillessen:2008qv}
S.~Gillessen, F.~Eisenhauer, S.~Trippe, T.~Alexander, R.~Genzel, F.~Martins and T.~Ott,
\emph{Monitoring stellar orbits around the Massive Black Hole in the Galactic Center},
\emph{Astrophys. J.} \textbf{692} (2009), 1075-1109
[arXiv:0810.4674 [astro-ph]].

\bibitem{Falcke:2013ola}
H.~Falcke and S.~B.~Markoff,
\emph{Toward the event horizon\textemdash{}the supermassive black hole in the Galactic Center},
\emph{Class. Quant. Grav.} \textbf{30} (2013), 244003
[arXiv:1311.1841 [astro-ph.HE]].

\bibitem{Reid:2014boa}
M.~J.~Reid, K.~M.~Menten, A.~Brunthaler, X.~W.~Zheng, T.~M.~Dame, Y.~Xu, Y.~Wu, B.~Zhang, A.~Sanna and M.~Sato, \textit{et al.}
\emph{Trigonometric Parallaxes of High Mass Star Forming Regions: the Structure and Kinematics of the Milky Way},
\emph{Astrophys. J.} \textbf{783} (2014), 130
[arXiv:1401.5377 [astro-ph.GA]].





\bibitem{Ishihara:2016vdc}
A.~Ishihara, Y.~Suzuki, T.~Ono, T.~Kitamura and H.~Asada,
\emph{Gravitational bending angle of light for finite distance and the Gauss-Bonnet theorem},
\emph{Phys. Rev. D} \textbf{94} (2016) no.8, 084015
[arXiv:1604.08308 [gr-qc]].

\bibitem{Crisnejo:2018uyn}
G.~Crisnejo and E.~Gallo,
\emph{Weak lensing in a plasma medium and gravitational deflection of massive particles using the Gauss-Bonnet theorem. A unified treatment},
\emph{Phys. Rev. D} \textbf{97} (2018) no.12, 124016
[arXiv:1804.05473 [gr-qc]].

\bibitem{Ovgun:2018fnk}
A.~\"Ovg\"un,
\emph{Light deflection by Damour-Solodukhin wormholes and Gauss-Bonnet theorem},
\emph{Phys. Rev. D} \textbf{98} (2018) no.4, 044033
[arXiv:1805.06296 [gr-qc]].

\bibitem{Ovgun:2018tua}
A.~\"Ovg\"un, \.I.~Sakall\i{} and J.~Saavedra,
\emph{Shadow cast and Deflection angle of Kerr-Newman-Kasuya spacetime},
\emph{JCAP} \textbf{10} (2018), 041
[arXiv:1807.00388 [gr-qc]].

\bibitem{Ono:2019hkw}
T.~Ono and H.~Asada,
\emph{The effects of finite distance on the gravitational deflection angle of light},
\emph{Universe} \textbf{5} (2019) no.11, 218
[arXiv:1906.02414 [gr-qc]].

\bibitem{Ovgun:2019wej}
A.~\"Ovg\"un,
\emph{Weak field deflection angle by regular black holes with cosmic strings using the Gauss-Bonnet theorem},
\emph{Phys. Rev. D} \textbf{99} (2019) no.10, 104075
[arXiv:1902.04411 [gr-qc]].

\bibitem{Javed:2019rrg}
W.~Javed, J.~Abbas and A.~\"Ovg\"un,
\emph{Effect of the Hair on Deflection Angle by Asymptotically Flat Black Holes in Einstein-Maxwell-Dilaton Theory},
\emph{Phys. Rev. D} \textbf{100} (2019) no.4, 044052
[arXiv:1908.05241 [gr-qc]].

\bibitem{Javed:2019ynm}
W.~Javed, R.~Babar and A.~\"Ovg\"un,
\emph{Effect of the dilaton field and plasma medium on deflection angle by black holes in Einstein-Maxwell-dilaton-axion theory},
\emph{Phys. Rev. D} \textbf{100} (2019) no.10, 104032
[arXiv:1910.11697 [gr-qc]].

\bibitem{Javed:2019kon}
W.~Javed, J.~Abbas and A.~\"Ovg\"un,
\emph{Deflection angle of photon from magnetized black hole and effect of nonlinear electrodynamics},
\emph{Eur. Phys. J. C} \textbf{79} (2019) no.8, 694
[arXiv:1908.09632 [physics.gen-ph]].

\bibitem{Zhu:2019ura}
T.~Zhu, Q.~Wu, M.~Jamil and K.~Jusufi,
\emph{Shadows and deflection angle of charged and slowly rotating black holes in Einstein-\AE{}ther theory},
\emph{Phys. Rev. D} \textbf{100} (2019) no.4, 044055
[arXiv:1906.05673 [gr-qc]].

\bibitem{Kumar:2019pjp}
R.~Kumar, S.~G.~Ghosh and A.~Wang,
\emph{Shadow cast and deflection of light by charged rotating regular black holes},
\emph{Phys. Rev. D} \textbf{100} (2019) no.12, 124024
[arXiv:1912.05154 [gr-qc]].

\bibitem{Li:2020ozr}
Z.~Li, Y.~Duan and J.~Jia,
\emph{Deflection of charged massive particles by a four-dimensional charged Einstein\textendash{}Gauss\textendash{}Bonnet black hole},
\emph{Class. Quant. Grav.} \textbf{39} (2022) no.1, 015002
[arXiv:2012.14226 [gr-qc]].

\bibitem{Pantig:2021zqe}
R.~C.~Pantig, P.~K.~Yu, E.~T.~Rodulfo and A.~\"Ovg\"un,
\emph{Shadow and weak deflection angle of extended uncertainty principle black hole surrounded with dark matter},
\emph{Annals Phys.} \textbf{436} (2022), 168722
[arXiv:2104.04304 [gr-qc]].

\bibitem{Uniyal:2022vdu}
A.~Uniyal, R.~C.~Pantig and A.~\"Ovg\"un,
\emph{Probing a non-linear electrodynamics black hole with thin accretion disk, shadow, and deflection angle with M87* and Sgr A* from EHT},
\emph{Phys. Dark Univ.} \textbf{40} (2023), 101178
[arXiv:2205.11072 [gr-qc]].

\bibitem{Pantig:2022toh}
R.~C.~Pantig and A.~\"Ovg\"un,
\emph{Dark matter effect on the weak deflection angle by black holes at the center of Milky Way and M87 galaxies},
\emph{Eur. Phys. J. C} \textbf{82} (2022) no.5, 391
[arXiv:2201.03365 [gr-qc]].

\bibitem{Kumaran:2022soh}
Y.~Kumaran and A.~\"Ovg\"un,
\emph{Deflection Angle and Shadow of the Reissner\textendash{}Nordstr\"om Black Hole with Higher-Order Magnetic Correction in Einstein-Nonlinear-Maxwell Fields},
\emph{Symmetry} \textbf{14} (2022) no.10, 2054
[arXiv:2210.00468 [gr-qc]].

\bibitem{Parbin:2023zik}
N.~Parbin, D.~J.~Gogoi and U.~D.~Goswami,
\emph{Weak gravitational lensing and shadow cast by rotating black holes in axionic Chern\textendash{}Simons theory},
\emph{Phys. Dark Univ.} \textbf{41} (2023), 101265
[arXiv:2305.09157 [gr-qc]].





\bibitem{Ishihara:2016sfv}
A.~Ishihara, Y.~Suzuki, T.~Ono and H.~Asada,
\emph{Finite-distance corrections to the gravitational bending angle of light in the strong deflection limit},
\emph{Phys. Rev. D} \textbf{95} (2017)  no.4, 044017
[arXiv:1612.04044 [gr-qc]].






\end{thebibliography}
\end{document}